\documentclass{emulateapj}

\shorttitle{A new abundance scale for 47\,Tuc}
\shortauthors{Koch \& McWilliam}

\slugcomment{To appear in the Astronomical Journal}

\begin{document}

\title{A new abundance scale for the globular cluster 47\,Tuc}

\author{Andreas Koch\altaffilmark{1} \& Andrew McWilliam\altaffilmark{1}}
\altaffiltext{1}{Observatories of the Carnegie Institution of Washington, 
                 Pasadena, CA, USA}
\email{akoch@ociw.edu; andy@ociw.edu}

\begin{abstract}
We present chemical abundances for O, Na, Mg, Al, Si, Ca, Ti and Fe 
in eight red giants and one turnoff star in the metal rich 
globular cluster 47\,Tuc, based on spectroscopy with
the MIKE high resolution spectrograph on the Magellan 
6.5-m Clay telescope.  
A robust line by line differential abundance analysis technique, relative to the K-giant Arcturus, 
was used to reduce systematic errors from atmospheric and atomic parameters. 
Our derived mean LTE [Fe/H] of $-0.76\pm0.01\pm0.04$ dex (random and systematic error respectively)
is more metal poor by about 0.1 dex than recent literature results. 
The chemical element ratios in this nearby globular cluster most closely resemble 
those of the Galactic bulge, although there is a non-negligible overlap with the composition of 
thick-disk stars.
We find that the [Al/Fe] and [Na/Fe] ratios coincide with the upper boundary of
the trends seen in the bulge and thick disk.  
There is only a small intrinsic scatter in the majority of the abundance ratios, indicating
that 47\,Tuc is mostly a rather chemically homogeneous system.  
\end{abstract}

\section{Introduction}
At a typical age range of 10--15\,Gyr, globular clusters constitute the oldest stellar objects in the 
universe and are thus ideal tracers of the early evolution of the Galaxy (Krauss \& Chaboyer 2003).  
For a profound understanding of their stellar populations, and their link to galactic evolution, 
it is essential to obtain accurate ages from photometric surveys. 
Presently, stellar ages are predominantly derived from isochrone fits to the red giant branch (RGB), 
the main sequence turnoff (MSTO) or white dwarf cooling sequences. 
However, the MSTO fit approach generally suffers from uncertainties in the distance modulus and 
unresolved reddening issues, whereas accurate observations toward the white dwarf regime 
are only sparsely available (e.g., Hansen et al. 2007). 
Furthermore, the chief impediment for obtaining accurate ages for stellar populations from RGB 
fits is the a priori unknown enhancement in $\alpha$-elements and the 
occurrence of age-metallicity degeneracies. 

The Space Interferometry Mission (SIM) will obtain accurate ($\sim$4 micro-arcsec) parallaxes to a
number of Population II objects (globular clusters and field stars in the halo) resulting in a
significant improvement in the Population II distance scale and greatly reducing the uncertainty
in the estimated ages of the oldest stars in our galaxy (Shao 2004; Chaboyer et al. 2005).
A critical component for improving the age determinations of globular clusters to 
$\sim$5\% is to measure absolute metallicities of the clusters with an accuracy of $\sim$0.05 dex.
The aforementioned problems are then resolvable if photometric and distance 
data are supplemented with 
accurate spectroscopic measurements of 
metallicities and chemical abundance ratios.  

Here we present a chemical abundance study based on a 
set of high resolution spectra of eight red giant stars and one subgiant in the
close (d=4.5 kpc; Harris 1996) Galactic globular cluster 47\,Tuc ($\equiv$NGC\,104). 
Previous work on 47\,Tuc that employed various stellar tracers, such as red giants
(Alves-Brito et al. 2005), subgiants and dwarfs (Carretta et al. 2004) and asymptotic giant
branch (AGB) stars (Wylie et al 2006), all agree on a mean metallicity near $-$0.7 dex; 
however, no consensus about an accurate absolute [Fe/H] has been reached, yet. 
In particular the age of 47\,Tuc has been the subject of debate: While 
Gratton et al. (2003) estimate an age of $\sim$11\,Gyr, thus 2.6\,Gyr younger than the old Galactic halo,  
McNamara (2001) and Liu \& Chaboyer (2001) suggest an age in excess of 12.5\,Gyr and thus 
comparable to the halo component. 
Several studies argue that 47\,Tuc's CMD 
bears striking resemblance with the solar metallicity bulge clusters 
NGC 6553 and NGC 6528 (Ortolani et al. 1995; Zoccali et al. 2003). In this light, 
refining 47\,Tuc's age and metallicity scale will also have a major impact on studies of 
the Galactic structure and the evolution of its components.  
Furthermore, the mean locus of 47\,Tuc's RGB, i.e., the cluster's photometric fiducials, 
has often been employed as a template to 
estimate the metallicity of resolved stellar systems such as 
dwarf spheroidal galaxies (e.g., Koch et al. 2006). Therefore, the reliable interpolation of a grid of 
fiducials requires a well defined metallicity scale (Sarajedini \& Layden 1997). 

Ideally, the effort of establishing a reliable age scale 
requires an absolute metallicity  scale with a desired 
accuracy of better than 0.05 dex in [Fe/H].
To achieve this, we perform our abundance analysis 
relative to the bright Galactic K-giant Arcturus, in order to minimize undesired dependences 
of the abundances on model atmospheres and atomic parameters, in particular log\,$gf$ values. 
In this, our work follows closely the recent bulge abundance studies of Fulbright et al. 
(2006;  2007, hereinafter F06 and F07). 

Finally, by means of the trends of the abundance ratios, one can approach the question, whether 
47\,Tuc can be plausibly assigned to any of the Galactic components. 
In the traditional division of Zinn (1985), 47\,Tuc is considered a typical disk cluster, 
which was later on specified by Armandroff (1989) in terms of an association with the {\em thick} disk. 
Dynamical evidence suggests, however, that this metal rich cluster rather exhibits bulge-like 
properties (Minniti 1995; C\^ot\'e 1999). 
Recent abundance results indicate that 47\,Tuc is dissimilar to either the typical disk, bulge 
or halo globular cluster populations. In this context, Alves-Brito et al. (2005) 
argue in favor of 47\,Tuc being 
a thick disk member. Common tracers of star formation as the $\alpha$-elements are found 
to be enhanced in this cluster, which would support the view of being an old system.
Clearly, there is no consensus about the status and membership of 47\,Tuc as yet. 

This paper is organized as follows: In \textsection 2 we present the data set and the standard reductions 
taken, while our atomic linelist and stellar atmospheres are briefly introduced in \textsection 3. 
The derivation of stellar parameters as well as a description of the abundance analysis relative to  
reference stars is presented in \textsection 4. Following the error analysis in \textsection 5, the 
metallicity scale and abundance results are  
discussed in \textsection 6. Finally, \textsection 7 summarizes our findings.

\section{Data \& Reduction}
Observations were carried out over four nights in July 2003 
using the Magellan Inamori Kyocera Echelle (MIKE) 
spectrograph at the 6.5m Magellan2/Clay Telescope (see the observing log in Table~1). 
The targets for this project were selected from the photometry of Kaluzny et al. (1998)
and infrared colors from the 2MASS database (Skrutskie et al. 2006).  
Our target list included eight RGB stars with V$-$K photometric temperatures within 200K of
the Arcturus temperature, plus one star near the cluster turnoff (TO) with a photometric temperature
close to the solar value (Figure 1).
A list of the targeted stars is given in Table~2, together with their  photometric properties. 
For the present data, we employed the full spectral coverage of the red echelle 
set up, which yielded a range of 4650--8300\AA. 
By using a slit width of 0.5$\arcsec$ and binning 2$\times$1 CCD pixels, we obtained a
spectral resolving power of R$\sim$40,000 for the RGB stars.  For the TO star we employed
a slit width of 0.75$\arcsec$ and 2$\times$2 binning, which gave a resolving power near
R$\sim$27,000.
Typically, each RGB star was observed for one hour, while for the TO star we obtained a
total exposure exceeding 11 hours, through variable sky conditions.  The observations were split
into several exposures to facilitate removal of cosmic ray effects. 

The pipeline reduction package of Kelson (2000; 2003) was used to reduce and extract 
the spectra from the raw data.
The extracted spectra were finally 
continuum normalized to first order by dividing through a blaze function. The latter was obtained 
by fitting a high-order polynomial to the high Signal-to-Noise (S/N) continuum spectrum of a
star from the very metal-poor cluster M30, which had
preferentially few, and very weak, absorption lines compared to the 47\,Tuc stars.
Typical S/N ratios, as defined by the peak value in the order containing 
the H$\alpha$-line, range from 100--150 per pixel for the bright red giants and reach $\sim$65 
for the fainter TO star. 

Radial velocities of the target stars were determined from the Doppler shifts of the order of ten 
strong, unsaturated and unblended absorption features. Variations of the individual 
stellar velocities are, however, not critical for our analysis, since our equivalent width measurement 
program, GETJOB (McWilliam et al. 1995a), determines the line center of each feature 
independently during our later measurement process. 
Overall, we find a mean radial velocity of $-15.4$ km\,s$^{-1}$\ with a dispersion of 6.0 km\,s$^{-1}$, 
where all of our 9 targeted stars are confirmed radial velocity members (see Table~2).

\section{Linelists}
For the purpose of our abundance analysis,  we measured equivalent widths (EWs) of a large number of 
absorption features using the semi-automated spectrum measurement routine 
GETJOB (McWilliam et al. 1995a). To first order, the lines were measured by an automated 
fit of a Gaussian line-profile, which then was manually improved by varying the spectral range to be included 
in the fit if necessary.  The average measurement uncertainty on the derived EWs, 
as determined from the r.m.s. scatter of the observed feature around the fit, is thus 
1.4\,m\AA, or typically less than 4\%.  Lines measured on the overlapping regions of 
adjacent orders generally agreed well to within this uncertainty and their final EW 
was then taken as the error weighted mean of the two values. 

In order to properly define the spectral ranges in which to perform the continuum fitting, 
we used the same, fixed continuum regions as in F06. These authors selected and fitted regions 
of a well-defined continuum by spectral synthesis of the metal rich red giant $\mu$\,Leo, 
accounting for the strong influence of line-blanketing. In doing so, we can yield 
a continuum setting accurate to  a few tenths of a per cent. 
The actual stellar continuum was then fit through these points by a polynomial of 
3rd to 4th order, such as to minimize the reduced $\chi^2$ of the fit. 

Our aim is to perform a differential line-to-line abundance analysis of our red giant 
sample relative to the well-studied K1.5 giant Arcturus ($\alpha$\,Boo). In doing so, 
our analysis becomes largely insensitive to the various shortcomings of model atmospheres, 
such as the use of plane-parallel, 2D and non-LTE approximations, but it also does not depend 
on the oft-uncertain atomic parameters, in particular the oscillator strengths of the absorption lines. 
The analysis of the TO star, on the other hand, requires another reference star that is 
preferentially close in its stellar parameters to those expected for the 47\,Tuc star, 
as estimated from its photometry and previously published metallicities. 
Hence, we chose to perform the differential analysis for our TO star 
relative to the metal-rich ([Fe/H]$\sim$$-0.64$ dex) Galactic field star Hip 66815  
(e..g., Fulbright 2000; Sobeck et al. 2006).  
Both Arcturus and Hip 66815 need then to be placed on the solar abundance scale, which 
we achieve by an analogous differential analysis, using the published solar EW measurements 
of Prochaska et al. (2000); F06, and F07.
Detailed spectrum synthesis of Arcturus and the Sun were employed by F06 to 
compile a line list that is essentially unaffected by contaminations from 
blended features.  
Moreover, the majority of these lines are sufficiently weak so that they still lie 
on linear part of curve of growth. 
As our work also relies on the analysis relative to Arcturus and since our target stars 
are similar in their properties to the bulge giant sample of F06, F07, we will 
use the linelist from F06 for the iron lines, and that from F07 for measuring 
the $\alpha$-elements in the following. In particular, we adopted the  EW measurements 
of F06, F07 for Arcturus to derive its log\,$\varepsilon$ values later on.
To further investigate potential biases from unresolved blends, F06 also 
calculated synthetic spectra for the super metal-rich  ([Fe/H]$\sim$+0.3 dex) field giant 
$\mu$\,Leo. From this, they constructed a ``clean'' linelist that is essentially free of 
blends, even at these high metallicities. Although neither Arcturus (at [Fe/H]=$-$0.5 dex; F06) 
nor any of our 47\,Tuc stars (at an expected [Fe/H] of approximately $-0.7$ 
dex according to previous estimates; e.g., Rutledge et al. 1997; Carretta et al. 2004, and 
references therein), are not expected to exhibit the comparable problem of blends as 
$\mu$\,Leo, we also ran our abundance analysis from only the clean linelist. 
However, as Figure 2 indicates, there is no significant difference seen when 
using the full linelist as opposed to only the clean subsample: the mean abundance ratio 
does not change upon variation of the linelist, and also the r.m.s. scatter of the mean
decreases by less than 0.01 dex. 

Many of the features in the red giant linelist will not be present in the (lower S/N) spectra of 
the hotter, high-gravity dwarf star. Thus, for our study of the TO star we supplemented the 
aforementioned linelist by suitable transitions from the thick disk study of Prochaska et al. (2000). 
The final linelists used throughout this work are given in Tables~3 and 4, together with the 
respective EW values in each of our targets. 

We note that we did not attempt to account for the Ca auto-ionization features (e.g., at 
6319\AA) in determining the \ion{Mg}{1} abundance and the affected weak lines were culled from the 
abundance determinations, whereas the stronger Mg line at 6318.75\AA\ yields an abundance 
estimate well in agreement with other, unaffected indicators.  
Finally, we did not include hyperfine splitting for the Na or Al lines in our analysis because the
corrections are negligible for these two elements. 

Finally, stellar abundances for each line were obtained using the {\em abfind} driver of 
the 2002 version of   MOOG (Sneden 1973).
The model atmospheres required for the abundance determinations will be 
detailed in the following Section. 

\section{Stellar atmospheres and parameters}

The stellar atmospheres to be used in the analysis were generated from the 
Kurucz non-LTE models\footnote{See \url{http://kurucz.harvard.edu} for the most recent version of these 
atmosphere grids.} without convective overshoot. As both Arcturus and 47\,Tuc 
show evidence of enhancement in the $\alpha$-elements by approximately $+0.4$ dex 
(e.g., Peterson et al. 1993; 
Carretta et al. 2004; F07) we used the  $\alpha$-enhanced opacity distributions 
AODFNEW\footnote{See \url{http://wwwuser.oat.ts.astro.it/castelli}} by F. Castelli, 
whereas the appropriate models for the Sun and Hip 66815 (with [$\alpha$/Fe]$<$0.2 dex) 
were calculated with the solar scaled distributions, ODFNEW. 

\subsection{Arcturus Revised}

For the Arcturus T$_{\rm eff}$\ we adopted the value of 4290\,K derived by F06 
from the angular diameter of Perrin et al. (1998) and the bolometric flux reported by 
Griffin \& Lynas-Gray (1999).  We note that the uncertainties on the bolometric flux, cited by 
Griffin \& Lynas-Gray led F06 to assign a total 1$\sigma$ T$_{\rm eff}$ uncertainty of only 10K.
We now believe that Griffin \& Lynas-Gray's flux uncertainty was too optimistic, as it is based
on essentially the same observational data as numerous other studies (Mozurkewich et al. 2003;
Alonso et al. 1999; Bell \& Gustafsson 1989; Augason et al. 1980),
yet those studies gave a range of bolometric flux values inconsistent with Griffin \& Lynas-Gray's 
stated flux uncertainty.  Here we adopt a 1 $\sigma$ uncertainty in the bolometric flux of 2.6\%
from the standard deviation of the fluxes estimated from the five investigations listed above.
This translates to a 1$\sigma$ T$_{\rm eff}$ uncertainty of 28K for Arcturus from the flux; 
the angular diameter contributes a temperature uncertainty of 8K, thus indicating a total temperature
uncertainty of 29K.

F06 estimated log\,$g$ = 1.6 and a microturbulent velocity parameter of 1.67 km\,s$^{-1}$ 
for Arcturus.  The Fe abundances in Arcturus, found using this initial 
set of stellar parameters, are shown in Figure 2. For this purpose, we consider the  
results relative to the Sun, for which we derived each line's abundance using the EWs
of F06, measured from the atlas of Kurucz et al. (1984) and the canonical solar
parameters of T$_{\rm eff}$=5777\,K, log\,$g$=4.44 and $\xi$=0.93 km\,s$^{-1}$.  A comparison of the 
Arcturus EWs measured from the Hinkle et al. (2000) Arcturus atlas with EW measurements of an Arcturus 
spectrum obtained with the MIKE spectrograph indicated that the MIKE EWs are $\sim$2\% larger; 
this corresponds to $\sim$0.008 dex in abundance, which is smaller than the computational 
uncertainties.

Figure~2 (top panel) shows that the run of abundance versus excitation potential
(henceforth ``excitation plot'') has essentially no slope, thus confirming the accuracy of the
adopted temperature scale.  The investigation  of  
the differences between abundance analyses using plane-parallel and spherical geometry model 
atmospheres by Heiter \& Eriksson (2006) predicts increased calculated abundances for low excitation 
lines of neutral iron-peak elements by up to $\sim$0.3 dex.  If the predictions 
are correct then the Arcturus excitation plot, using solar gf values, should show a slope when the 
correct temperature is adopted for Arcturus.  However,  Figure~2, 
indicates that, empirically, the absolute effect for Arcturus is rather small ($\sim$0.05 dex), 
and therefore, the differential effect between Arcturus and the 47~Tuc stars must be negligible.
This supports the validity of the excitation plots
to constrain the temperatures of the 47~Tuc red giants.
The same arguments apply to the possible effects of granulation on the derived abundances
of \ion{Fe}{1} lines as a function of excitation potential (see Steffen \& Holweger 2002): Figure~2 shows
that the difference in this effect is not detected between the Sun and Arcturus and so should be 
negligible when taken differentially between the 47~Tuc giants and Arcturus.

Our initial data showed, however,  a considerable trend of iron abundance in Arcturus with EWs. This prompted us 
to lower the microturbulence to 1.54 km\,s$^{-1}$\ in order to remove any such remaining slope (lower panel of Figure~2). 
One noteworthy feature in these plots is, however, the lack of ionization equilibrium: 
While the mean iron abundance, [Fe/H], from the neutral stage is $-0.49 \pm 0.01$ dex and thus consistent with 
the value of $-0.50$  from Peterson et al. (1993) and  F06, 
the ionized species yield an abundance [\ion{Fe}{2}/H] higher by 0.08 $\pm 0.03$ dex.   

Our measured deficiency of \ion{Fe}{1}, relative to \ion{Fe}{2}, is similar to the 0.03 to 0.06 dex non-LTE
effect predicted for the K0III star Pollux by Steenbock (1985): the non-LTE correction for \ion{Fe}{1}
lines of high excitation potential were $\sim$0.02 dex, while low excitation lines required corrections
of $\sim$0.05 dex, and the \ion{Fe}{2}
correction was approximately $-$0.01 dex.  Thus, a
non-LTE excitation temperature for Pollux would be lower than the LTE value (a trend of increasing 
correction with line EW was also predicted, but in an LTE 
abundance analysis this would be absorbed into the microturbulent velocity parameter). 
For iron in the Sun Steenbock (1985) found non-LTE corrections of 0.01 and 0.03 dex for weak high and low
excitation \ion{Fe}{1} lines, respectively.  More recent non-LTE calculations by Korn, Shi \& Gehren (2003) 
found that collisions of hydrogen atoms with iron in the solar atmosphere are enhanced over the Darwin (1968, 1969) 
formalism by a factor of three (S$_H$=3); since Steenbock (1985) employed the Darwin formulae 
(i.e., S$_H$=1) the Steenbock's non-LTE corrections for the sun and Pollux are probably overestimated.
Pollux is 560K hotter than Arcturus (McWilliam 1990) and more metal rich by  $\sim$0.4 dex, so the non-LTE
corrections may differ.  However, F07 showed that the \ion{Fe}{1}/\ion{Fe}{2}  
and \ion{Ti}{1}/\ion{Ti}{2} abundance ratios for bulge and disk red giants are approximately constant 
in the [Fe/H] interval from $-$1.5 to $+$0.5 dex and for 
T$_{\rm eff}$ from 4000 to 5000K; this suggests that
the non-LTE corrections for Pollux and Arcturus are probably very similar.  
If we assume similar non-LTE correction for Arcturus and Pollux for an equal mix of low and high excitation
lines, and if we include the solar non-LTE corrections from Steenbock (1985) and Korn, Shi \& Gehren (2003), 
then our estimate for the non-LTE corrected Arcturus [Fe/H] from the neutral lines is $-$0.49 dex.  The
non-LTE corrected \ion{Fe}{2} lines indicate [Fe/H] of $-$0.46; this iron ionization difference, at
0.03 dex, is comparable to the 1$\sigma$ random error on the mean \ion{Fe}{2} abundance, derived from a mere
5 lines.

In order to explore the effect of gravity of iron ionization equilibrium
we opted to re-derive the mass of Arcturus by comparing its location in the 
T$_{\rm eff}$-luminosity diagram with theoretical isochrones. The luminosity was 
obtained from its {\em Hipparcos}-catalog 
based apparent magnitude and distance of ($-0.04\pm0.02$)\,mag, ($11.25\pm0.09$)\,pc, 
respectively.  A bolometric correction (BC) of ($-0.78\pm0.02$) was adopted from an interpolation 
of the predicted photometry from the Kurucz-atmosphere grid.  For consistency 
it is also necessary to employ the predicted BCs with the Kurucz-based bolometric 
magnitude for the Sun, at 4.62, when computing stellar photometric gravities.  
The BCs implied by the Teramo  group (``BaSTI''; Pietrinferni et al. 2004)  
could have been used with the standard M$_{\rm bol}$ of 4.74
for the Sun (e.g., Cox 2000) to give the same result.  Unfortunately,
F06; F07 did not appreciate this problem, so their computed gravity for
Arcturus was too small by 0.04 dex.  

We employed the solar composition and $\alpha$-enhanced BaSTI
isochrones (Pietrinferni et al. 2004; including the revised $\alpha$-opacities from Ferguson 
et al. 2005) to estimate the mass and age of Arcturus, as indicated by its location in the
T$_{\rm eff}$ versus M$_V$ plane.  Our analysis involved use of cubic spline 
interpolation of the BaSTI grid to the metallicity, Z, and $\epsilon$(O/Fe) of Arcturus, 
taken from the abundances of F07.  In particular we interpolated to the Arcturus Z value 
(of 0.00954) for solar composition and $\alpha$-enhanced BaSTI isochrones, followed by 
interpolation to the Arcturus $\epsilon$(O/Fe) value of 1.67 dex.  It is
important to note that although the $\alpha$-enhancements of Mg, Si and Ca employed in the
BaSTI models are similar to the values seen in Arcturus, the O/Fe value of the models is
0.2 dex higher than observed; presumably this resulted from early high estimates of the 
solar oxygen abundance (e.g. Grevesse \& Noels 1993) used in the BaSTI models.

Our results indicate a best fit age of 9.0$^{+3.0}_{-1.8}$ Gyr and a mass of 1.00$^{+0.05}_{-0.09}$ 
 M$_{\odot}$ for Arcturus.  The uncertainties on age and mass include estimates 
for the uncertainties of T$_{\rm eff}$, distance, magnitude, metallicity and O/Fe ratio.

The BaSTI results are well consistent with the value derived from the $\alpha$-enhanced 
Padova isochrones (Salasnich et al. 2000), from which we obtain (0.92$\pm$0.09)\,M$_{\odot}$
and  (10.5$\pm$1.0)\,Gyr.  Lastly, we employed the Yonsei-Yale (Yi \& Demarque 2003) set of
isochrones, again accounting for enhanced $\alpha$-element abundances of  [$\alpha$/Fe]=+0.4 dex. 
As Maraston (2005) argues, the BaSTI models tend to produce higher T$_{\rm eff}$\ for a given
luminosity and mass compared to the other sets of isochrones. Hence the latter will necessarily
result in lower age estimates. Concordant with this view, the Yonsei-Yale tracks provide a
``best-fit'' mass of  (1.22$\pm$0.2)\,M$_{\odot}$ and a corresponding younger age of  
(5$\pm$1.5)\,Gyr.  In the following, we will use the BaSTI based mass estimate, since these
tracks incorporate a sophisticated, i.e., varying with metallicity, treatment of the mixing
length and have succeeded best in matching the observations of Galactic globular clusters 
(Maraston 2005).  Higher adopted mass, and accordingly log$\,g$, would aggravate the departure 
from ionization equilibrium, whereas  coincidence of the neutral and ionized stage would 
require log$\,g\sim1.44$, which is established with a lower mass of 0.71\,M$_{\odot}$.  

If we appeal to errors in the basic parameters of $\alpha$\,Boo to lower its mass, 
 then an increase in T$_{\rm eff}$ of $\sim$190K, or an
increase in the metallicity, Z, by 0.2 dex are required; these changes are far greater than
the uncertainty in these two parameters, and so can be eliminated.  Another possibility is that
Arcturus has experienced mass-loss, perhaps related to its metallicity and evolutionary
phase, or as the result of interactions with a putative companion (e.g. Soderhjelm \& Mignard 1998;
Verhoelst et al. 2005; but see Griffin 1998).  While this extra-ordinary mass loss is feasible, 
and may reflect lack of certainty of mass loss in metal-rich RGB stars, it seems a rather 
ad-hoc assumption.  Also, it remains unlikely that the observed temperature and luminosity 
are consistent with a lower mass of Arcturus.

We note in passing that the isochrone mass-loss parameter is not sufficient to reduce the
derived Arcturus mass enough to obtain Fe ionization equilibrium.
At the location of Arcturus on the RGB the mass-loss parameter values, $\eta$, of 0.2 and 0.4
(Reimers 1977) hardly affect the stellar mass; however, at the tip of the RGB and on the AGB 
these parameters do have a significant effect.  Although it would enable ionization equilibrium
the luminosity and temperature of Arcturus eliminates it as a possible AGB star.
In the following, we adopt a mass of 1.00 M$_{\odot}$ for Arcturus, based on the BaSTI 
isochrones, which corresponds to a gravity of 1.64 $\pm$0.03. 

\subsection{47\,Tuc  -- The Red Giant Sample}

The T$_{\rm eff}$\ values of the target stars were calibrated using the empirical color-temperature 
relations from Alonso et al. (1999).  For this purpose, we exploited the 
V and I photometry from Kaluzny (1998), complemented by Infrared J and K data from the 
2MASS (Skrutskie et al. 2006).  The 2MASS photometry was converted to the TCS system, 
employed in the Alonso et al. (1999) T$_{\rm eff}$\ calibration, using the Carpenter (2005 unpublished) 
2MASS--CIT transformation\footnote{\url{http://www.astro.caltech.edu/$\sim$jmc/2mass/v3/transformations/}} 
and the CIT--TCS conversion of Alonso et al. (1998).

In addition to the color transformations it is necessary to correct for interstellar extinction;
we employ the extinction law of Winkler (1997) throughout this work.
Harris (1996) obtained a reddening for 47~Tuc of E(B$-$V)=0.05 mag, while the reddening maps
of Schlegel et al. (1998) indicate E(B$-$V)=0.032 mag.  Gratton et al. (2003) obtained
E(B$-$V)=0.035 $\pm$0.009 from B$-$V colors and E(B$-$V)=0.021 $\pm$0.005 from {\em b$-$y}
colors.  The average of all four estimates is 0.035 $\pm$0.006; if the unusually
low reddening from {\em b$-$y} colors by Gratton et al. (2003) is ignored, the average reddening 
becomes 0.039 $\pm$0.007 mag.  Considering this distribution of reddening estimates for 47~Tuc
for our initial reddening value we adopt E(B$-$V)=0.04$\pm$0.01 mag.  

The large baseline of the V$-$K color and the steep color-temperature gradient for red giant
stars makes this T$_{\rm eff}$\ estimate the least susceptible to potential photometric uncertainties.
As a result temperatures derived from V$-$K color are generally expected to be the most 
accurate and we shall use
these values as the photometric T$_{\rm eff}$\ in the following. The typical uncertainty on V$-$K temperatures,
given by the photometry and calibration errors, is $\sim$40\,K.

We chose to employ the Alonso et al. (1999) (V$-$K) color temperatures differentially, 
relative to the known (V$-$K) and T$_{\rm eff}$ of Arcturus, in order to remove possible 
zero-point systematics in the calibration.  The average of photometry from Lee (1970) and
Johnson et al. (1966) were used to determine the Arcturus V$-$K 
color of 2.925 mag (in the Johnson-Cousins system); we note that
the V$-$K color from the two studies differ from the mean by 0.025 mag.
The Johnson V$-$K color, when converted to the CIT system using the relations in Elias et al. (1982)
and then to the TCS system, as described above, indicate T$_{\rm eff}$ = 4252K using the Alonso et al. (1999)
calibration; this is 38K cooler than the directly measured effective temperature of 4290$\pm$29K, 
based on absolute flux and angular diameter measurements.  Thus, in order to place our 47~Tuc photometric
V$-$K temperatures onto the physical T$_{\rm eff}$\ scale for Arcturus it is necessary to add a zero point of 38K to
the values indicated by the Alonso et al. (1999) V$-$K calibration. In this way
our photometric T$_{\rm eff}$ values are differential to the Arcturus value.  We note that the
$\pm$0.025 mag range of the observed (V$-$K) Arcturus color corresponds to 17K.

In addition to the V$-$K photometric temperatures we derived spectroscopic temperatures by demanding
excitation equilibrium for Fe, i.e. that there is no slope in the plot of iron abundance versus 
excitation potential.  Since each individual 47~Tuc line abundance is differential to the
same line in Arcturus we ensured that our spectroscopically determined temperatures are on the
same physical T$_{\rm eff}$\ scale as Arcturus.
The differential V$-$K photometric and spectroscopic temperatures in this work are listed in Table~5.
The mean difference between the two temperatures, T(V$-$K)$-$T(spec), is $-$5K with an r.m.s. scatter of 41K; this 
suggests a random error on the mean difference from the 8 RGB stars of only $\pm$15K.  We note that a reddening
change of 0.01 mag. in E(B$-$V) leads to a $\sim$19K change in photometric T$_{\rm eff}$; thus, the agreement
between our photometric and spectroscopic T$_{\rm eff}$\ values is consistent with the range of reddening values
cited in the literature, from 0.03 to 0.04 mag.

We adopted physical gravities for the 47~Tuc stars based on Equation~1:   
\begin{equation}
\log{g} = \log{M/M_{\odot}} - \log{L/L_{\odot}} + 4\log{(T/T_{\odot}}) + \log{g_{\odot}}
\end{equation}
For the luminosities we used
the dereddened  V-band photometry of Kaluzny (1998), BC from the Kurucz web site and
a distance modulus of (m$-$M)$_0$=13.22$\pm$0.06.  Our adopted distance modulus is the average from six studies
(Zoccali et al. 2001; Percival et al. 2002; Gratton et al. 2003; Weldrake et al. 2004; 
McLaughlin et al. 2006; Thompson et al. 2008 in preparation) that used numerous techniques, 
including HST proper motions, the white dwarf cooling sequence, empirical CMD fitting, and eclipsing binaries.

Masses for the 47~Tuc stars were determined by comparison with the latest version of the Teramo 
$\alpha$-enhanced isochrones (with $\alpha$ element opacities from Ferguson et al. 2005).  In Figure~1
we compare the observed 47~Tuc color-magnitude diagram with the $\alpha$-enhanced, 
[Fe/H]=$-$0.70 dex, Teramo 
12 Gyr isochrone; this is the average of the ages estimated by Zoccali et al.  (2001) and Percival et al. 
(2002).  This comparison suggests a typical mass of 0.89M$_{\odot}$ for our 47~Tuc RGB stars.  

For a more precise comparison of the isochrones with the 47~Tuc giants we interpolated the isochrones 
to match the overall metallicity, Z, and the measured O/Fe ratio from a previous iteration on the 
abundances (Z=0.0062, $\epsilon$(O/Fe)=1.77).  The observed M$_V$ values of our 
47~Tuc RGB stars were closest to the interpolated isochrones with ages in the range 12 to 14 Gyr.  However, these
interpolated isochrones were all brighter than the observed M$_V$ values by a few hundredths of a magnitude; 
although, the differences are well within the systematic uncertainties.  For example: a T$_{\rm eff}$ increase of 10 to 30K, 
or a reddening increase of 0.01 mag (slightly more than the uncertainty on the observed Arcturus V 
magnitude) could account for the difference.  This range of isochrone ages gave masses (for $\eta$=0.4) of 
0.87 M$_{\odot}$ for the 12 Gyr isochrone to 0.81 M$_{\odot}$ for 14 Gyr.

These isochrone-based masses compare favorably with the direct measurement of 0.871$\pm$0.006 M$_{\odot}$  for 
a subgiant in a double-lined, eclipsing, spectroscopic binary in 47~Tuc by Thompson et al. (2008).   Inspection of 
the 12 Gyr, $\alpha$-enhanced,  Teramo isochrones shows that the masses of the subgiant and our RGB
stars should be equal to within 0.01 M$_{\odot}$.  This is the case because the mass loss experienced by the RGB
stars reduces their slightly higher original mass to the mass of the subgiant, according to the $\eta$=0.4
prescription of the Teramo isochrones.  Given the consistency of these methods we adopt the direct measurement of
Thompson et al.  (2008), at 0.87 M$_{\odot}$ for the mass of our 47~Tuc RGB stars.
Since star \#3 is closer to the AGB track in the 
CMDs shown in Figure~1 we adopt the AGB mass of 0.65M$_{\odot}$ for this star, as indicated by the interpolated
Teramo isochrones, interpolated to the Z and O/Fe measured from the spectra.
In the abundance iterations the overall metallicity parameter, [M/H], for the input model atmospheres,
is equated to [\ion{Fe}{1}/H] abundance from the previous iteration. 

As an initial guess for the microturbulence parameter, $\xi$, for the model atmospheres, we used the 
same value as in Arcturus, due to the expected similarity in the spectral type of our 
target stars and $\alpha$\,Boo. This value of 1.54 km\,s$^{-1}$ provides a reasonable 
estimate for 4 of our 8 stars in terms of no apparent trend of \ion{Fe}{1} abundance with EW. 
The weakest lines with EWs below 12\,m\AA\ were neglected for this part 
of the analysis.  
Nevertheless, stars \#6--\#8 required a $\xi$ lower by $\sim0.1$ km\,s$^{-1}$, while 
for star \#3, this parameter had to be significantly increased (see Figure~3) to eliminate any such slope,
as might be expected from its AGB star evolutionary status. 
Usually, $\xi$ could be determined to within $\pm0.1$ km\,s$^{-1}$ accuracy. 

Ionization equilibrium is not obtained for any of our eight red giants in 47~Tuc;
the mean [\ion{Fe}{1}/\ion{Fe}{2}] ratio is $+$0.083 $\pm$0.015 dex with a scatter of
the 8 stars of $\sigma$=0.040 dex.
The giant that comes closest to $\epsilon$(\ion{Fe}{1})=$\epsilon$(\ion{Fe}{2}) is star \#7
with [\ion{Fe}{1}/\ion{Fe}{2}]=$+$0.03 dex. This issue will be addressed in 
detail in Sect.~4.4. 

The final chemical abundance ratios derived for the 47~Tuc red giant sample, using the 
spectroscopic excitation 
T$_{\rm eff}$\ values are listed in Table~6. 

\subsection{47\,Tuc -- the turnoff star}
The atmospheric parameters of the unevolved 47~Tuc TO star differ substantially from those
of the red giant sample, so Arcturus can no longer be employed as a reference for differential
abundance analysis. 
The 47~Tuc TO star was selected with T$_{\rm eff}$\ similar to the Sun, in order to perform a solar 
differential abundance analysis.  However, due to its low metallicity lines in the TO star are
considerably weaker than the same lines in the Sun.  As a result of the low $S/N$ of the TO star
spectrum lines that are measurable in the TO star are saturated in the sun, thus
reducing the effectiveness of a differential abundance analysis.

To solve this this difficulty we employed a two-step differential abundance analysis using
a bright Galactic halo field TO star, Hip\,66815.  Its 
parameters (5850K, logg=4.5, $\xi$=0.95 km\,s$^{-1}$, [Fe/H]=$-$0.64; [Fulbright 2000; Sobeck et al. 2006]) 
are comparable to those expected for the 47\,Tuc TO star, judging from its location in color-magnitude
space and the previously discussed red giant sample.  
Our $S/N\sim$500 spectrum of Hip\,66815 enabled us to measure very weak lines,
down to $\sim$5 m\AA , in this star that are not saturated in the solar spectrum.  Thus, we 
used unsaturated lines in the 47~Tuc TO star and Hip\,66815 for the first step, and then compared
very weak, but well measured, lines in Hip\,66815 to unsaturated lines in the solar spectrum
for the second step of the differential analysis.
These lines and their EWs are listed in Table~4.  

We refined the stellar parameters of Hip\,66815 and the 47\,Tuc TO star in
the same manner as discussed above for the red giant sample. 
Thus it turned out that the temperature of Hip\,66815 needed to be lowered by 100\,K from the
photometric value.
Contrary to the red giants, damping issues such as Stark and Van der Waals 
broadening play a progressively dominant role in the dense atmospheres of the dwarf stars. 
 For the case of hydrogen we matched the
observed H$\alpha$-line profile with a spectral synthesis, in which the damping constants in the classical Uns\"old
approximation were multiplied by a constant factor of 6.5, which provided the best fit to the line wings. 
In this case we obtained an independent, reddening-free estimate of the stellar T$_{\rm eff}$ by 
synthesizing the temperature sensitive wings of the H$\alpha$ line.  From this we found a T$_{\rm eff}$
value in agreement with the photometric temperature to within 40\,K.
In the following we employed this  damping factor enhancement of 6.5 over the Uns\"old approximation
for all the absorption lines. It should be noted 
that errors due to this approximate treatment will be absorbed by
a change in the best fit microturbulence parameter. 

In this way we found a [Fe/H] of $-$0.76 for the reference star Hip\,66815, 
which is about 0.1 dex more metal poor than derived in earlier studies 
(Fulbright 2000; Sobeck et al. 2006); it should be noted that the Sobeck et al. (2006) work assumed
the Fulbright T$_{\rm eff}$\ value.  Fulbright's excitation T$_{\rm eff}$\ was based on laboratory
log\,$gf$ values, which may constitute a significant source of uncertainty, whereas our differential 
analysis is largely insensitive to the systematic effects induced by such atomic parameters. 
The photometric and excitation temperature of the 47\,Tuc TO star  
agree remarkably well to within $\la$20\,K. 

Ionization equilibrium is well established in Hip\,66815, 
it does, however, not hold in the 47~Tuc TO star. We address this issue 
in detail in the next Section. 
Our final abundance ratios for the 47~Tuc TO star are presented in Table~7.

\subsection{Problems with Ionization Equilibrium in 47~Tuc}

As can be seen from Table~6 the mean difference between \ion{Fe}{1} and \ion{Fe}{2} 
abundances ($\Delta$Fe=$\epsilon$(\ion{Fe}{1})$-$$\epsilon$(\ion{Fe}{2})) for our 47~Tuc giants
(relative to Arcturus) is 0.08 dex.  In this Section we investigate possible resolutions of this 
apparent inconsistency. We note that the differential [\ion{Fe}{1}/H]
and [\ion{Fe}{2}/H] values were computed based on an assumed [Fe/H] for Arcturus of $-$0.51 dex,
from the \ion{Fe}{1} lines alone. 
It is interesting that if we employ our
Arcturus [\ion{Fe}{2}/H] value to compute the differential [\ion{Fe}{2}/H] for the 47~Tuc giants, and
\ion{Fe}{1} lines to compute [\ion{Fe}{1}/H], the ionization equilibrium problem is reduced to only 
$\Delta$Fe=0.03 dex; i.e. consistent to within 1$\sigma$ of satisfying ionization equilibrium 
for Fe.   Thus, it appears that the differential abundances of 47~Tuc relative to the Sun 
provides ionization equilibrium, while iron ionization equilibrium for Arcturus is not satisfied.

We remind the reader that the purpose of employing results differential to Arcturus was to
avoid systematic errors in the model atmospheres, due to errors and the use of incomplete
physical description in the 1D Kurucz model grid.  
Hence, it is also the difference between the
Arcturus and 47~Tuc atmosphere parameters that determine whether ionization and
excitation equilibrium are achieved.
To explain the ionization imbalance it is then necessary to find an unaccounted systematic
difference between the Arcturus and 47~Tuc red giant atmospheres.  

In the mean,  ionization equilibrium could be achieved
by an average increase in log~$g$ for the 47~Tuc giants of $\sim$0.16 dex. 
As indicated by eq.~1, mass, luminosity and temperature determine the gravity.
If a systematic error in log~$g$ is the source of the apparent under-ionization of Fe,
the change in log~$g$ cannot be due to mass, as this would indicate the presence
of main sequence stars with M$\sim$1.4 M$_{\odot}$ (late F-type), which are not seen
in 47~Tuc.
For luminosity to create this change in log~$g$ would require a 0.5 magnitude
decrease in the distance modulus, implying a distance to 47~Tuc closer by 500 pc, or $(m-M)_0$=12.72. 
This value lies $8\sigma$ below the
mean value, and 0.3 magnitudes lower than the lowest recent measurement of the distance modulus.
Uncertainties in the photometry are less than a few hundredths of a magnitude, given the consistency 
of the Kaluzny (1998) CMD with other published work (e.g., Armandroff 1988).  We 
estimate the uncertainty on the Kurucz BCs at 0.05 magnitudes, based on
the uncertainties in temperature, gravity and metallicity. 
Finally, the published reddening values for 47~Tuc cite uncertainties of 0.01 magnitudes.
The quadrature sum of these uncertainties affecting the luminosity amount to 
0.04 dex in log~$g$, much less than the 0.16 dex in log~$g$ required to explain the apparent
under-ionization of iron.

At first look it may seem that
the required changes in log~$g$ cannot be due to temperature, either, given the excellent agreement
of the two sets of independent differential temperatures for our red giants
on the Arcturus scale; it would be necessary to investigate how both temperature measures 
could be in error.  However, the Fe ionization imbalance could be resolved with a modest 
systematic T$_{\rm eff}$ decrease of $\sim$60K, as can also be seen from Table~8. 
This would, however, lower the \ion{Fe}{1} abundances by only $\sim$0.01 dex, and 
fortunately does not affect our conclusion on the [Fe/H] scale of  our 47~Tuc red giants.
 The main
effect is the enhanced temperature sensitivity of the \ion{Fe}{2} abundance, most likely 
due to a rapid change in the degree of Fe ionization at the T$_{\rm eff}$ of Arcturus.
Because of the differential nature of both the photometric and spectroscopic temperatures, relative 
to Arcturus, a shift in the adopted Arcturus T$_{\rm eff}$ does not affect the relative derived
temperatures of the 47~Tuc giants; thus, the ionization equilibrium is not directly affected by any
reasonable change in the Arcturus temperature.  
 
Next, we investigated the sensitivity of ionization equilibrium to the adopted model atmosphere
[$\alpha$/Fe] ratio.  An increase from scaled-solar composition to the 0.4 dex enhancement
of the Castelli/Kurucz $\alpha$-enhanced model atmospheres raised the computed 
\ion{Fe}{2}/\ion{Fe}{1} abundance ratio by $\sim$0.10 dex.  This suggests that models with
[$\alpha$/Fe] enhancements of $\sim$0.8 dex for 47~Tuc would solve the Fe ionization 
problem -- a value that has not been detected that highly in any globular cluster so far.  
Inspection of the abundances in Table~6 shows average $\alpha$ element 
(O, Mg, Si, Ca, Ti) enhancement of 0.41 dex.  In this regard it is interesting that the 
average $\alpha$-enhancement for Arcturus is 0.34 dex; if Na and Al are included then the 
47~tuc stars are enhanced by 0.39 dex and Arcturus is enhanced by 0.31 dex.  To estimate 
the effect of the difference in composition of the light metals between Arcturus and 47~Tuc
we computed the electron densities (e.g. see Mihalas 1978) for the Arcturus and 47~Tuc 
compositions at the T$_{\rm eff}$\ of star \#8, with gas pressures indicated by the scaled
solar and $\alpha$-enhanced models.  The change in electron density due to the composition
difference was about one third of the change from scaled-solar to $\alpha$-enhanced models,
which suggests that the composition difference could account for $\sim$0.03 dex of the 
difference between \ion{Fe}{1} and \ion{Fe}{2} abundances in the 47~Tuc stars.  Thus, with
the correction for composition difference between Arcturus and the 47~Tuc giants we only
need to explain an ionization difference of 0.068 dex, which corresponds to a temperature 
decrease of $\sim$40K.

If we adopt the Lee (1970) V$-$K for Arcturus, at 2.90 mag, rather than the average of the Lee
(1970) and Johnson et al. (1966) results, the 47~Tuc V$-$K colors, differential to 
Arcturus, appear cooler, by 16K.  Next, if we employ a reddening for 47~Tuc of 
E(B$-$V)=0.027 mag, instead of the previously adopted 0.040, then our 47~Tuc 
photometric temperatures are reduced by an additional 24K.  This is not an unreasonable 
assumption: if the notably high reddening value of Harris (1996) is omitted, the unweighted
average of the Schlegel et al. (1998) value and two the values from Gratton et al. (2003) 
gives E(B$-$V)=0.029 $\pm$0.005.
Unfortunately, we cannot find a satisfactory fit to the observed CMD with E(B$-$V)=0.02
and the standard theoretical isochrones of the Teramo group: in order to fit the RGB an older age
is required with the lower reddening, but this then creates a poor fit for the core helium burning,
red clump, stars.  On the other hand a reddening of E(B$-$V)=0.04 gave a good fit
to the observed CMD (see Figure~1).  

If the differential photometric V$-$K temperatures are in fact  cooler by 50--60K we
need to understand why the excitation of \ion{Fe}{1} lines indicates a higher temperature.
Because our excitation analysis is differential to the Arcturus line abundances, the
excitation temperatures are themselves differential to the adopted Arcturus temperature;
thus, if we lower the Arcturus T$_{\rm eff}$ value the excitation temperatures will
decline accordingly.
One possibility
is that there are systematic measurement differences between the Arcturus
spectrum and the 47~Tuc spectra, since we did not obtain a spectrum of Arcturus with the
same instrument as the 47~Tuc spectra; rather, we employed EWs measured from the high
$S/N$, high resolving power Hinkle et al. (2000) Arcturus atlas.  

Another possibility is that 
the Kurucz models approximate the 47~Tuc stars differently than Arcturus,
resulting from real physical differences between Arcturus and 47~Tuc stars.  
The difference between \ion{Fe}{2} and \ion{Fe}{1} abundances could be understood
if Arcturus is reddened by E(B$-$V) $\sim$0.03 magnitudes, or if
Arcturus is an AGB star, and thus less massive than our CMD fit indicated.
However, there is no evidence for this amount of reddening, and the V magnitude
of Arcturus is too faint, by 0.15 magnitudes, for it to be an AGB star.  These
possibilities would affect the derived excitation temperatures, but not the differential
photometric temperatures.

It is also  not possible to correct the inconsistency between \ion{Fe}{1} and \ion{Fe}{2} abundances
in 47~Tuc by altering the helium mass fraction, Y, of the cluster, because it would be necessary to 
drastically reduce 
Y --  a reduction from Y=0.20 to an unrealistically low value of Y=0.10 changes the gravity by less
than 0.1 dex (Green, Demarque \& King 1987). Not only would
such a low Y value be inconsistent with Big Bang nucleosynthesis theory, but it is still insufficient
to resolve the observed abundance differences in our 47~Tuc giants.  
On the other hand,  an appealing alternative is that the helium fraction of Arcturus is significantly higher
than normal expectations.  Helium-rich Teramo isochrones, with Y=0.40, were 
kindly supplied to us by Santi Cassisi (2008, private communication). These isochrones fit the 
Arcturus temperature, luminosity and composition with an age of 14 Gyr and a mass of 
0.67 M$_{\odot}$.
This old age offers a natural explanation for the enhanced [$\alpha$/Fe] ratio in
Arcturus, which would be difficult to understand if it is $\sim$3 Gyr younger than the globular
clusters, as indicated from the isochrones with standard He composition (see Sect.~4.2), 
because supernovae (SNe) of Type~Ia would be expected to have reduced the [$\alpha$/Fe] ratio over  
that time. 
An increase in the He fraction of $\alpha$\,Boo would, however,  result in an opposite problem for 
the (differential) \ion{Fe}{1} and \ion{Fe}{2} abundances for the bulge and disk giant 
stars in F06 and F07. Thus, it seems that the helium mass fraction cannot be altered to resolve
the Fe ionization problem in our 47~Tuc red giants.
Also, if Arcturus has log~g near 1.44 dex, consistent with the Y=0.40 tracks, 
then both its [\ion{Fe}{1}/H] and [\ion{Fe}{2}/H] values decrease to $-$0.55 dex, 
while the mean of the [Fe/H] values for our 47~Tuc
RGB stars remains unchanged, at $-$0.76 dex (see also Sect. 6).   

It is possible that differential non-LTE effects, between Arcturus and the 47~Tuc giants,
could result in over-estimated excitation temperatures of the 47~Tuc giants (see Sect. 4.1). 
Steenbock (1985) predict that the non-LTE corrections for low excitation \ion{Fe}{1} lines in 
Pollux are larger than for the high excitation \ion{Fe}{1} lines by about 0.03 dex. 
From our LTE calculations we estimate that the resulting excitation temperature difference would be
of order 50K.  Thus, in order to obtain the over-estimated excitation temperatures in
47~Tuc giants, relative to Arcturus, it is necessary for the non-LTE abundance effect
on \ion{Fe}{1} in the 47~Tuc giants to be larger than Arcturus by about a factor of $\sim$2. 
Since 47~Tuc giants are more metal-poor and slightly more luminous than Arcturus 
the 47~Tuc giants have more ionizing photons and fewer collisional de-excitations in the line
forming region than in Arcturus. Therefore, one might expect slightly greater non-LTE effects
in the 47~Tuc giants, in the right sense to ultimately understand the LTE iron ionization 
equilibrium problem encountered here. 

It is interesting to note that Ti ionization equilibrium does hold for our 47~Tuc stars,
with a mean deviation $\Delta\epsilon$(\ion{Ti}{1}$-$\ion{Ti}{2}) of $0.00\pm0.02$ dex. 
However, for the coolest 47~Tuc red giant in our sample, the
\ion{Ti}{1} abundance is higher than \ion{Ti}{2} by 0.08 dex (1.6\,$\sigma$). 
If we lower the adopted T$_{\rm eff}$\ of our 47~Tuc giants by $\sim$50K then the Ti~I$-$Ti~II
abundance difference will decrease to $-$0.11 dex.  Thus, we are unable to obtain ionization
equilibrium for both Fe and Ti simultaneously, although a deficiency of Ti~I relative to Ti~II
might readily be understood as a consequence of non-LTE over-ionization of Ti.
 A more precise estimate of the effect of
non-LTE on differential excitation temperatures in red giants, however,  
requires a detailed non-LTE study, 
which is beyond the scope of the current paper.

Finally, we note that the ionized species in the 47~Tuc TO star gives an abundance higher by 0.14 dex than
the neutral one, while ionization equilibrium is well established in the reference dwarf star Hip 66815. 
Because of the similarity of \ion{Fe}{1} and \ion{Fe}{2}, which deviate by a mere 0.02 dex, in the reference star,
which has similar parameters to the 47~Tuc TO star, non-LTE can be eliminated as a possible explanation
for the ionization imbalance in the 47~Tuc TO star.  
Th\'evenin \& Idiart (1999) estimate that for subgiants and TO stars with T$_{\rm eff}$\ and log\,g 
similar to our sample a non-LTE treatment of the iron lines yields a [Fe/H] higher by $\sim$0.1 
dex, which is  close to the order of magnitude of the discrepancy seen in our star. 
However,  Th\'evenin \& Idiart (1999) did not include line opacities in their
non-LTE calculations, which magnified the predicted non-LTE effects.
We suggest that either the measured \ion{Fe}{2} lines in the 47~Tuc TO star are affected by noise
spikes that bring them into detection in the low S/N spectrum, or that the electron density in this star
is higher than in Hip\,66815, perhaps due to an unusual $\alpha$ enhancement.

\section{Abundance error estimates}
In order to quantify the sensitivity of our derived abundances to the uncertainties in the 
atmospheric parameters,  we performed a standard error analysis.  For this purpose, we
computed nine different atmosphere models including the following, conservative 
changes in atmosphere
parameters: T$_{\rm eff}$ $\pm50$\,K,  log\,$g\,\pm\,$0.2 dex (which is of the order of 
the maximum departure from ionization equilibrium observed in our stars),
$\xi\pm0.1$ km\,s$^{-1}$, and the input metallicity [M/H] by $\pm$0.1\,dex. We also
included calculations with solar [$\alpha$/Fe] ratios, using the Kurucz ODFNEW model 
atmospheres.  We then derived abundances in the same manner as described in Sections 4.2 and 4.3 for
the 47~Tuc TO star and the two red giant stars in our sample that cover the largest 
possible range of T$_{\rm eff}$\ and log\,$g$ (stars \#1 and \#8).  The resulting abundance deviations 
are illustrated in Table~8.

As can be seen from Table~8, our mean \ion{Fe}{1} abundance values are largely insensitive to 
the choice of T$_{\rm eff}$.  On the other hand, changes of log\,$g$ mainly upsets \ion{Fe}{2} and leaves 
the neutral stage unscathed as is expected due to the strong gravity dependence of the ionization. 
Furthermore, there is only a negligible influence of the microturbulence and overall metallicity 
on the mean abundances.  

In practice, the $\sim$40K r.m.s. dispersion between 47~Tuc giant excitation T$_{\rm eff}$\ values and
V$-$K color T$_{\rm eff}$\ values (Sect. 4.2) 
suggests a random $1\sigma$ uncertainty for each T$_{\rm eff}$\ 
indicator near 30K.  Combining this in quadrature with the systematic T$_{\rm eff}$\ uncertainty
for $\alpha$\,Boo of the same order, we obtain a total T$_{\rm eff}$\ uncertainty of 42K.

The distance modulus uncertainty, at 0.06 mag, dominates the uncertainty in the 47~Tuc
giant  M$_{\rm v}$ values.  The combination of $\delta$M$_{\rm v}$ and $\delta$T$_{\rm eff}$\ 
values results in typical mass uncertainties of 0.05 M$_{\odot}$, from placement of the
giants onto the best fit BaSTI isochrone.
For the stellar luminosities we include a BC uncertainty of 
0.05\,mag, based on the dependence of BC on atmosphere parameters, a V-band magnitude
uncertainty of 0.02 mag. and the distance modulus uncertainty of 0.06 mag.  When these
are combined in quadrature we obtained a total uncertainty in log~L/L$_{\odot}$ of 0.032 dex.
Finally, in the gravity calculation of Equation~1 we adopted the total T$_{\rm eff}$\ uncertainty of
42K, the luminosity uncertainty of 0.032 dex and the mass uncertainty of 0.05 M$_{\odot}$
to obtain a 1$\sigma$ uncertainty in log~g of 0.043 dex; conservatively we adopt a value of
$\pm$0.05 dex for log~g.

To determine the net effect of these actual parameter uncertainties on the final abundances we
combine in quadrature the contributions from T$_{\rm eff}$\ and log~g above, determined from Table 8, 
together with a 0.04 km/s
uncertainty on $\xi$, based on the scatter of abundances in the EW versus $\epsilon$(Fe)
plot, a 0.05 dex uncertainty in the model atmosphere [M/H], and a 0.1 dex uncertainty in [$\alpha$/Fe]. The latter 
corresponds to 1/4 of the abundance difference resulting from use of ODFNEW vs. AODFNEW model
atmospheres.  As a result, we find a total 1$\sigma$ systematic abundance uncertainty of 0.04 dex
for [\ion{Fe}{1}/H] and 0.08 dex for [\ion{Fe}{2}/H].
One should keep in mind, however, that all atmospheric parameters are covariant to a certain extent
(McWilliam et al. 1995b) so that our stated abundance uncertainties are upper limits.   

We investigated the effect of the use of MARCS\footnote{We obtained the MARCS atmosphere grid from
\url{http://marcs.astro.uu.se/}} models (Gustafsson et al. 2003) on our derived abundances.  The giant
atmosphere models were computed with spherical geometry, so the abundances should be computed using
this assumption; however, 2-D radiative transfer it is not currently available with MOOG.  Therefore, in
order to facilitate use of MOOG for the abundance calculations we generated 1-D equivalent models from the
2-D MARCS grid.  We found rather small differential abundance results, with
 $\Delta$$\epsilon$(Kurucz)$-$$\Delta$$\epsilon$(MARCS)=0.03 dex for both Fe~I and Fe~II lines.  This minor
abundance systematic would reduce the final abundances of our 47~Tuc stars if MARCS models were employed. 

Listed in Tables~6,7 are also  the statistical error in terms of the 1\,$\sigma$ scatter of our 
measurements from individual lines and the numbers of lines employed to derive these
abundances. 
In the case, where there is a sufficient amount of measurable 
lines in the spectra, i.e., for \ion{Fe}{1}, the random error on the mean abundance is small compared
to the one induced by the  atmospheric uncertainties.  Conversely, if only a handful of lines is present,
such as for \ion{Fe}{2}, this statistical error will dominate.  We find the random r.m.s. abundance scatter 
in the range 0.03 to 0.12 dex, or typically $\sim$0.07 dex per line.

\section{Abundance results}

\subsection{A new metallicity scale for 47\,Tuc}

From our entire sample of nine stars we find a mean iron abundance of 
[\ion{Fe}{1}/H]  = $-0.76\pm0.01\pm0.04$, where the first value gives the internal error, i.e., 
the standard deviation of the mean and the second number incorporates the systematic 
errors.  At 0.03\,dex, the star-to-star scatter within this 
cluster is very small and fully consistent with our estimated measurement errors, 
as is also seen in the recent spectroscopic 
data of Carretta et al. (2004) and  Alves-Brito et al. (2005) as well 
in the photometric estimates of  Hesser et al. (1987). 
In particular, our stars cover a range in [\ion{Fe}{1}/H] from $-0.82$ to $-0.72$, 
Moreover, it is reassuring that the TO star yields an iron abundance that is 
is excellent agreement with the red giant sample, indicating that the chemical iron 
abundance in this cluster is independent of the stars' evolutionary status. 

Apart from the early work of Brown \& Wallerstein (1992), who derived a lower iron 
abundance of  $-0.81$\,dex, all previous studies have placed  47\,Tuc  towards 
the more metal rich regime. For instance, Alves-Brito et al. (2005) find a [Fe/H] 
of ($-0.66\pm0.12$)\,dex, and Kraft \& Ivans (2003) found [\ion{Fe}{2}/H]=$-$0.79, based on
an empirical calibration of Ca-triplet metallicities, but from a re-analysis of both
the Brown \& Wallerstein (1992) and Caretta \& Gratton (1997) EWs they obtained $-$0.70 dex,
based on corrections for their preferred models and log~$gf$ values.
Moreover, the recent AGB star study of Wylie et al. (2006) argues in favor of the metal richer 
regime in terms of [Fe/H]=($-0.60\pm0.20$)\,dex. 
While these estimates can still be considered to coincide with our value to within the quoted 
errorbars, it is more difficult to reconcile the result of  ($-0.67\pm0.04$)\,dex by 
Carretta et al. (2004) from their accurate analysis of 12 subgiant and dwarf stars 
with our data.  Recently, McWilliam \& Bernstein (2008) have obtained [\ion{Fe}{1}/H]=$-$0.75$\pm$0.03
dex based on an integrated-light analysis of the core of 47~Tuc. While the agreement with the
present result is encouraging the analysis method was radically different, except that
McWilliam \& Bernstein also performed a differential analysis relative to $\alpha$~Boo.

The main reason for discrepancies between our results and other studies can be 
sought in the different analysis method employed.  Virtually all the previous works 
performed their analyses relying on the accuracy of the oscillator strength. 
Despite the availability of accurate laboratory log\,$gf$ values (e.g., Blackwell 
et al. 1995), the use of these values in abundance studies of red giants may be fraught with 
peril (e.g., Prochaska et al. 2000), in particular, when the derived abundances are 
referenced to the solar abundance scale. 
We sidestepped such difficulties by performing our line-by-line differential 
analysis relative to Arcturus so that  unavoidable systematic uncertainties, which  may 
affect the different approaches, can be expected to have canceled out to first order. 
Examples of effects on standard abundance analysis that may, to first order, cancel out in
our differential technique include errors in $gf$ values, non-LTE effects, spherical versus
plane parallel atmosphere geometry, inhomogeneous photospheres (granulation), 3-dimensional hydrodynamics and the effect
of chromospheres.

One might argue that all the studies under scrutiny have focused on 
stars in different evolutionary stages, reaching from the unevolved early subgiants and dwarfs 
in the Carretta et al (2004) sample to AGB stars in the case of Wylie et al. (2006) and 
evolved red giants in the data of Alves Brito et al. (2005). However, all these works agree 
in that they do not find any considerable abundance spread with evolutionary status, 
and also we cannot confirm a significant difference between our red giant sample 
and the TO star's abundance ratios. Moreover, also our likely AGB star candidate (\#3) 
is entirely consistent with the remainder of our iron results. Hence we conclude that 
the difference between our results and the previous works is 
not due to any target selection effects.

\subsection{Alpha-elements: O, Mg, Si, Ca, Ti}

An overview of our resultant abundance ratios is provided in the Box-Whisker plot in Fig.~4, which 
displays each element's mean, the 25- and 75\% interquartile ranges as well as the full range 
covered.  It turns out that each of Si, Ca and Ti are similarly enhanced at $\sim$0.37\,dex\footnote{In 
order to gain a better agreement of their thin disk stars with recent results from disk dwarf and
TO stars, F07 shifted their [Si/Fe] abundance ratios by $-$0.09 dex. 
In the following, we will also apply this offset to {\em their} values in our diagrams.}. 
On the other hand, there is an indication that Mg shows even higher abundances for most of the stars, 
with an average of [Mg/Fe]$\sim$0.45\,dex.   Oxygen is also enhanced more than Si, Ca and Ti, in
our 47~Tuc stars, with a mean [O/Fe] value, at $+$0.60 dex.
These results are in excellent agreement with the abundances determined in the bulge giants
of F07 with similar [Fe/H], as indicated by the filled circles in Fig.~4. 
Overall, it is worth noticing that {\em all} of the [$\alpha$/Fe] abundance ratios only show 
very little star-to-star variations, where the scatter does not exceed 0.06\,dex for 
the elements considered here and can be purely explained by the random EW measurement 
errors of our analysis.  

The abundance ratios for the TO star are consistent with those in the 
red giant sample, with the exception of [Mg/Fe], where the TO star yields an abundance 
lower by 0.12\,dex, and for the case of [Si/Fe], with an abundance higher by 0.07\,dex in the 
dwarf.  
We note that Si and Mg in dwarf stars are both sensitive to damping and the adopted stellar
gravity; an overestimate of gravity would result in over-abundances of these two elements.
The question of why these two species show opposite deviations remains open, 
but it may indicate overestimated EWs for Si, due either to blends or noise effects (see also Cohen et al. 2004).

The production of the  $\alpha$-elements is generally believed to occur in SNe II, 
which are associated with the death of massive, short-lived stars. 
However, while Mg and O are believed to be be formed during the hydrostatic nuclear burning phase 
in the SNe II progenitors, 
Si, Ca and Ti are generally assigned to the explosive nucleosynthetic phase of 
the SNe II (e.g., Woosley \& Weaver 1995). It is thus natural to expect different trends 
of for instance [Mg/Fe] as compared to [Si,Ca,Ti/Fe] as a function of [Fe/H]. 
Following F07, we investigate, to what extent these groups of $\alpha$-elements track each other's  
trends: we plot in Fig.~5 the difference of Ca and Si, Ca and \ion{Ti}{1}, 
respectively,  with metallicity.  It is reassuring that all of the explosive $\alpha$-elements 
follow the identical trends with [Fe/H], which is reflected in that the aforementioned abundance ratios 
exhibit flat slopes with metallicity near solar values. At 0.04 ([Ca/Si]) and 0.05 dex ([Ca/\ion{Ti}{1}]), 
the scatter around zero is small.   

Again following F07 we consider the
straight average [$<$SiCa\ion{Ti}{1}$>$/Fe] as a measure of explosive $\alpha$-element abundance 
to reduce the overall scatter in these elements.  
Fig.~6 shows this average abundance ratio in comparison with the bulge data of F07. 
Furthermore, this diagram incorporates a number of abundance ratios from the  Galactic thin disk 
(Reddy et al. 2003;  Bensby et al. 2005; Brewer \& Carney 2006) 
and the thick disk (Prochaska et al. 2000; Fulbright 2000; Bensby et al. 2005; Brewer \& Carney 2006). 
 
As was shown by F07 the Galactic 
bulge is clearly separated from the thin disk in the $\alpha$-elements by $\sim +$0.2 dex, and 
the bulge [$\alpha$/Fe] ratio shows a distinct decline with increasing metallicity.  This picture 
is generally interpreted in terms of a higher SNe II / SNe Ia ratio in the bulge, indicating an increased 
importance of massive stars for the evolution of the bulge, and addition of iron from long-lived SNe~Ia
at higher [Fe/H] than for the thin disk.
The Galactic thick disk $\alpha$-abundances, on the other hand, are significantly lower than 
the bulge values at [Fe/H]$\sim -$0.5, while these two components converge  
towards the enhanced [$\alpha$/Fe] ratio at metallicities around $-1$. 
How does now 47\,Tuc fit into this picture? As Fig.~6 implies, this cluster is clearly 
strongly enhanced in its $\alpha$-elements and inconsistent with an association 
with the thin disk. 
At a metallicity of [Fe/H]=$-$0.76, the cluster stars occupy a region in [$\alpha$/Fe] that
is higher than most thick-disk stars, and consistent with the bulge composition. And yet, at this 
[Fe/H],  the bulge [$<$SiCa\ion{Ti}{1}$>$/Fe] value is higher than the thick disk population by only 
$\sim$0.1 dex.  Thus, while the {\em mean} 47~Tuc [$<$SiCa\ion{Ti}{1}$>$/Fe] ratio is similar to the
bulge, there is a small overlap with the composition of thick-disk stars.

The fact that [Mg/Fe] in the bulge is higher on average than the explosive $\alpha$-elements' 
abundance ratios was first noted by McWilliam \& Rich (1994) and later confirmed 
by F07.  This enhancement as well as only a negligible decline of [Mg/Fe] with [Fe/H] 
was interpreted by a rapid formation of the bulge (Matteucci et al. 1999).
Also [Mg/Fe] in the 47\,Tuc stars shows an excess of $\sim +0.1$\,dex relative to the [Si,Ca,Ti/Fe] 
group (Fig.~4 and bottom panel of Fig.~7), while [O/Fe] has an excess of $\sim$0.2 dex.  This is in
accord with the conclusion of F07 that
explosive and hydrostatic $\alpha$-elements show different trends in the bulge, as is also seen in
the dwarf galaxy abundances (see the review by Venn et al. 2004).  As mentioned in F07 part of the reason 
for the lower enhancement of the [Si,Ca,Ti/Fe] relative to the hydrostatic alphas is that SNe~Ia are 
thought to produce small amounts of Si,Ca and Ti; so that ratios taken relative to the solar composition 
are lower because of the SNe~Ia contribution of these elements in the sun.
 
\subsection{Aluminum and Sodium}
The first obvious thing to note is that both Na and Al are enhanced (by +0.2 and +0.45 dex, respectively)  
in 47 Tuc. While the [Na/Fe] enhancement is in good agreement with that found 
in previous studies (Brown \& Wallerstein 1992; Norris \& Da Costa 1995; Carretta et al. 2004; 
James et al. 2004; Alves-Brito et al. 2005), these works do not support evidence for 
an average [Al/Fe] in excess of 0.38\,dex, with the exception of Brown \& Wallerstein, who 
yield an estimate of +0.67. 
Generally, a wide range of [Al/Fe] ratios for various stellar systems is not uncommon 
and globular cluster giants have been reported to reach maximum enhancements in excess of 
$+1$\,dex (McWilliam 1997; Ivans et al. 1999). 

The top panel of Fig.~7 shows our derived [Al/Fe] abundances; note that the raw [Al/Fe]
bulge abundances of F07 are used the comparison.  In a similar plot F07 shifted their [Al/Fe] ratios
lower to account for zero-point differences between studies of disk dwarfs and giants.  However, as indicated
in F07, non-LTE effects on the [Al/Fe] ratios in the dwarf studies are probably to blame for the shift,
so the dwarf results should have been corrected upward, although the amount depends on the stellar temperatures.
The F07 bulge giant [Al/Fe] ratios should have remained unaltered.  Thus, in the top panel of Fig.~7 the 
thin and thick disk results, which were based on dwarf stars, should be increased by an average of 0.08 dex.
Figure~7 shows that our mean [Na/Fe] and [Al/Fe] ratios for 47~Tuc are at the upper end of the distributions 
found in the bulge, but also compatible with the, few, thick disk stars in this metallicity regime.  

Although both the light elements Na and Al 
are also produced by SNe II, the trends with metallicity are clearly distinguished from those 
found for the other $\alpha$-elements discussed in the previous section (see F07 for a review). 
While the Galactic thin and thick disk abundance ratios exhibit a smooth but slow decline 
with increasing [Fe/H], which is due to the continuous pollution by SNe II and a late addition 
of SNe Ia processed material into the ISM, [Al/Fe] in the bulge shows the opposite trend. 
This can be understood in terms of the production of significantly less iron in the Galactic bulge 
environment.  

Given the same line of reasoning as for the traditional enhancement of Na and Al in the 
Galactic bulge, we can argue that the material, out of which 47 Tuc formed had lesser iron
contributions from SNe\,Ia, but more important, the Al and Na yields are slightly more increased 
at the metallicity of 47\,Tuc.  This could have occurred as a result of an overly high neutron
excess (Arnett 1971) or an increased fraction of high mass stars in the population that 
produced the 47~Tuc composition.  In this context Vuissoz et al. (2004) suggest that rotating
Wolf-Rayet stars can provide increased Al yields.  However, the presence of material from
more massive SNe~II events should result in higher [O/Mg] ratios in 47~Tuc than the bulge, which 
we do not observe.  

Again, the lowest [Al/Fe] is found for the TO star and the AGB-star candidate \#3. 
Non-LTE calculations for Al in dwarf stars by Baum\"uller \& Gehren 
(1997), indicate a correction of $+$0.09 dex for the \ion{Al}{1} lines we used in the TO 
star. If this correction is applied the TO star, the [Al/Fe] ratio increases to 
$+$0.42 dex, similar to the mean value of $+$0.47 for the 47~Tuc giants. 
Similar calculations for Na (Baum\"uller, Butler \& Gehren 1998;  
Takeda et al. 2003), indicate downward corrections to the TO star Na 
abundance of approximately 0.06 dex.  While these non-LTE corrections 
bring the TO star and giants into better agreement it must be remembered 
that, in order to obtain absolute abundance ratios, non-LTE treatment needs 
to be considered for all element species, including Fe. This holds for the TO star, 
the RGB stars and the sun, but such a treatment is beyond the scope of the current paper.

We note in passing that there is no clear evidence of the Mg-Al anticorrelation (Langer \& Hoffmann 
1995) present in our data, which would be indicative of the occurrence of deep mixing 
processes  in globular cluster stars. 
This finding is also consistent with the result of Carretta et al. (2004). 

On the other hand,  we cannot exclude the possibility that Na and O anticorrelate in 47\,Tuc (Fig.~8).  
Such a trend was suggested to occur also in  this metal rich cluster (e.g., Carretta et al. 2004), 
and to be present across all stellar evolutionary stages. The canonical picture to explain this 
abundance anomaly is the occurrence of internal deep mixing during the first dredge-up.  
Our data covers only a small range in the [O/Fe] abundances, but 
it is clear that the star with the highest depletion in oxygen is accordingly enhanced in 
Na by about 0.15 dex above average. 
All in all, there is a large scatter present in 
the [Na/O] abundance ratio (at an r.m.s. of 0.11 dex) and the object with the highest enhancement 
in [Na/Fe] appears to fall into the thin disk regime. Moreover, in the phase space of sodium vs. oxygen 
(Fig.~8, bottom) the over-abundance of sodium coincides with the upper 
end of the bulge and thick disk abundances, which is, however, hampered by the sparse sampling of 
thick disk stars in this regime. The general enhancement of Na in 47\,Tuc can 
then also be understood, if, as for the case of the Galactic bulge, there was a primordial 
source for the Na production available (F07).

\section{Summary \& Discussion}

In this work we have determined the chemical abundance ratios of 8 red giants and one TO star 
in the Galactic globular cluster 47\,Tuc. By focusing on the derivation of the O, Na, Mg, Al, Si, Ca and 
Ti abundances, we could investigate the properties of those elements usually associated with 
their production in the massive SNe II events.  
One major improvement of the present work compared to previous studies is the 
differential line-by-line abundance analysis relative to a reference star with similar temperature,
gravity and metallicity
as the target stars. By thus referencing our measurements to the red giant Arcturus and 
a field dwarf, we could efficiently reduce undesired uncertainties in the stellar model 
atmospheres and prevent potentially erratic atomic parameters, thereby 
achieving a high degree of accuracy in our abundance results.  
By means of detailed isochrone fits, we estimated Arcturus' mass and age 
to be 1.0\,M$_{\odot}$ and 9\,Gyr, confirming its nature as a first ascent red giant branch star. 
Our derived  mean LTE [Fe/H] of $-$0.76 $\pm$0.01 $\pm$0.04 is more metal-poor by about 0.10 dex
than recent results in the literature.
This new accurate measurement will enable future age determinations of this globular cluster 
to an unprecedented accuracy and  can also be employed to establish the new metallicity 
scale with 47\,Tuc as a template of metal rich cluster. 

We confirmed that the explosive $\alpha$-elements Si, Ca and Ti trace each other's abundance 
trends, which lends support to their common production site in the explosive phase of the SNe\,II 
events. Magnesium, however is enhanced relative to these elements.  We find that [Al/Fe] and
[Na/Fe] ratios in 47~Tuc lie on the upper bound of the distribution seen in Galactic bulge stars
of similar metallicity.  Our data show a mild evidence of a Na-O anticorrelation in that the stars
with lower [O/Fe] are slightly enhanced by 0.1 dex in [Na/Fe]. Comparable relations of Mg with Al
could not be confirmed.   

All of the abundance trends we found are inconsistent with those seen in the Galactic thin disk, which 
suggests that 47\,Tuc, like the Galactic bulge and thick disk, was presumably 
subject to a rapid formation with very little contributions from the long-lived SNe\,Ia. 
There is no {\em a priori} reason to assume that this cluster should bear any association with 
either the bulge or the thick disk component. 
On the other hand, 
most of chemical abundance trends indicate 
that 47\,Tuc exhibits  
bulge-like properties, such as the overall $\alpha$-enhancement, its over-abundance 
of Mg and in particular the mild enhancement in [Na/Fe] at the stars' [O/Fe] metallicities.
If  47\,Tuc in fact formed from bulge material, it must have been 
subject to rapid formation period and accompanying enrichment. 
It is worth noting that the scatter in essentially all the abundance ratios
is quite small and most likely solely due to the random EW measurement errors. 
This points to the cluster having formed and evolved in a homogeneous environment. 

One drawback of the bulge-scenario are the cluster's present day large Galactocentric 
distance ($R_{GC}=7.4$\,kpc; Harris 1996) and its height below the plane ($Z=3.2$\,kpc) 
that makes a correlation with the bulge less likely. Moreover, its 
pericenter at 5.2\,kpc and its low eccentricity orbit ($e\sim0.16$; Dinescu et al. 1999) 
render it unlikely that its origin has been close to the Galactic bulge, whereas an initial  
thick disk connection cannot be ruled out at present.
In this context, 
the heavy elements measured by Alves-Brito (2005) suggest that 
47\,Tuc cannot be unequivocally assigned to either of the thin disk, bulge or halo globular cluster 
population as established by Zinn (1985), but rather should be associated with the 
Galactic thick disk. In fact, also our overall abundance results could sustain this view, 
since the average [X/Fe] ratios we measured partially overlap with those from the thick disk. 
 
The formation of the thick disk is nowadays believed to be related to a dissipationless 
merger event that heated the early thin disk (Walker et al. 1996; Prochaska et al. 2000; 
Gilmore et al. 2002; Feltzing et al. 2003;  Wyse et al. 2006).  
In this context it is possible that 47\,Tuc formed out of the material that later on 
constituted the heated disk. According to Bensby et al. (2004), thick disk stars in the metallicity 
regime of 47\,Tuc exhibit typical ages older than 11\,Gyr, which is also consistent 
47\,Tuc's age of $\ga$11\,Gyr (e.g., Liu \& Chaboyer; Grundahl et al. 2002; Gratton et al. 2003). 
Alternatively, the cluster could have formed from material that was stripped from 
the infalling satellite and subsequently dispersed into the newly formed disk, or  
it might have already been in place in the infalling satellite and then been tidally stripped. 
Yet some of the abundance trends that we find defy a simple allocation  
of 47\,Tuc to the thick disk; for instance, the high [Al/Fe] ratio in the globular cluster are at odds with 
the significant enrichment by SNe\,Ia in the disk as reported, e.g.,  by Prochaska et al. (2000); 
Feltzing et al. (2003), and  
at this stage we refrain from over-interpreting any association due to the sparse sampling 
of thick disk stars around 47\,Tuc's metallicity in the literature. 

\acknowledgments
We gratefully acknowledge funding for this work from a NASA-SIM key project grant, entitled ``Anchoring 
the Population II Distance Scale: Accurate Ages for Globular Clusters and Field Halo Stars''.
This research has made use of the NASA/ IPAC Infrared Science Archive, which is operated by the Jet
Propulsion Laboratory, California Institute of Technology, under contract with the National Aeronautics
and Space Administration

\clearpage

\begin{figure}[!ht]
\begin{center}
\includegraphics[angle=0,width=0.4\hsize]{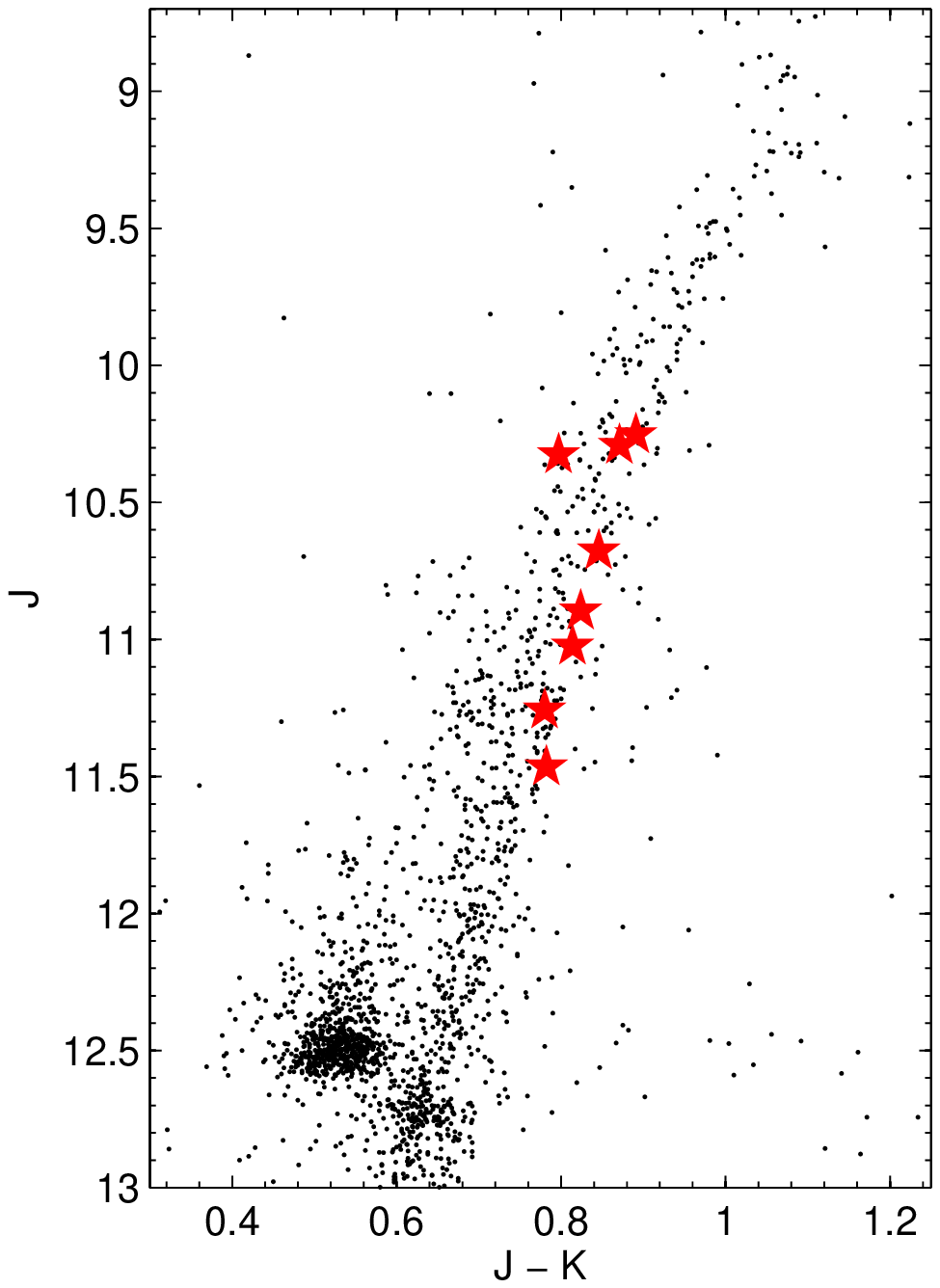}
\includegraphics[angle=0,width=0.6\hsize]{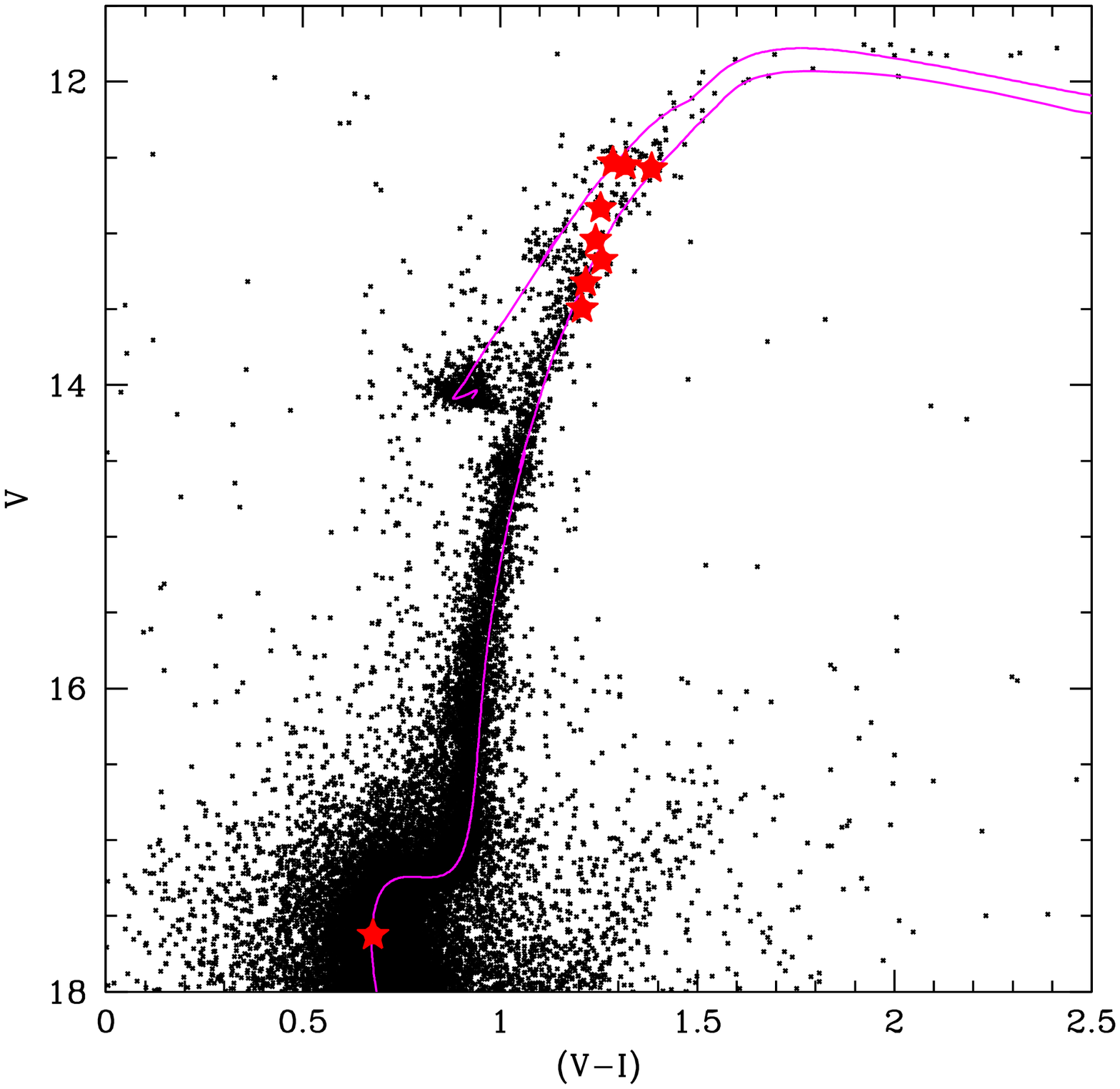}
\end{center}
\caption{Location of our target stars in color-magnitude space. The J and K magnitudes 
were taken from the 2MASS survey (Skrutskie et al. 2006), whereas the V and I photometry is 
that of Kaluzny et al. (1998). The dwarf star (at V$\sim$17.6) is below the magnitude limit 
of 2MASS.  The observations in the lower panel are compared to a 12 Gyr alpha-enhanced, $\eta$=0.4, theoretical
Teramo isochrone with [Fe/H]=$-$0.70; the isochrone is shifted for E(B$-$V)=0.03, 
(m$-$M)=13.22 and a (V$-$I) color shift of $-$0.033 mag to put the Teramo colors on the 
Alonso et al. (1999) T$_{\rm eff}$--(V$-$I) system. }
\end{figure}
\begin{figure}[!ht]
\begin{center}
\includegraphics[angle=0,width=0.6\hsize]{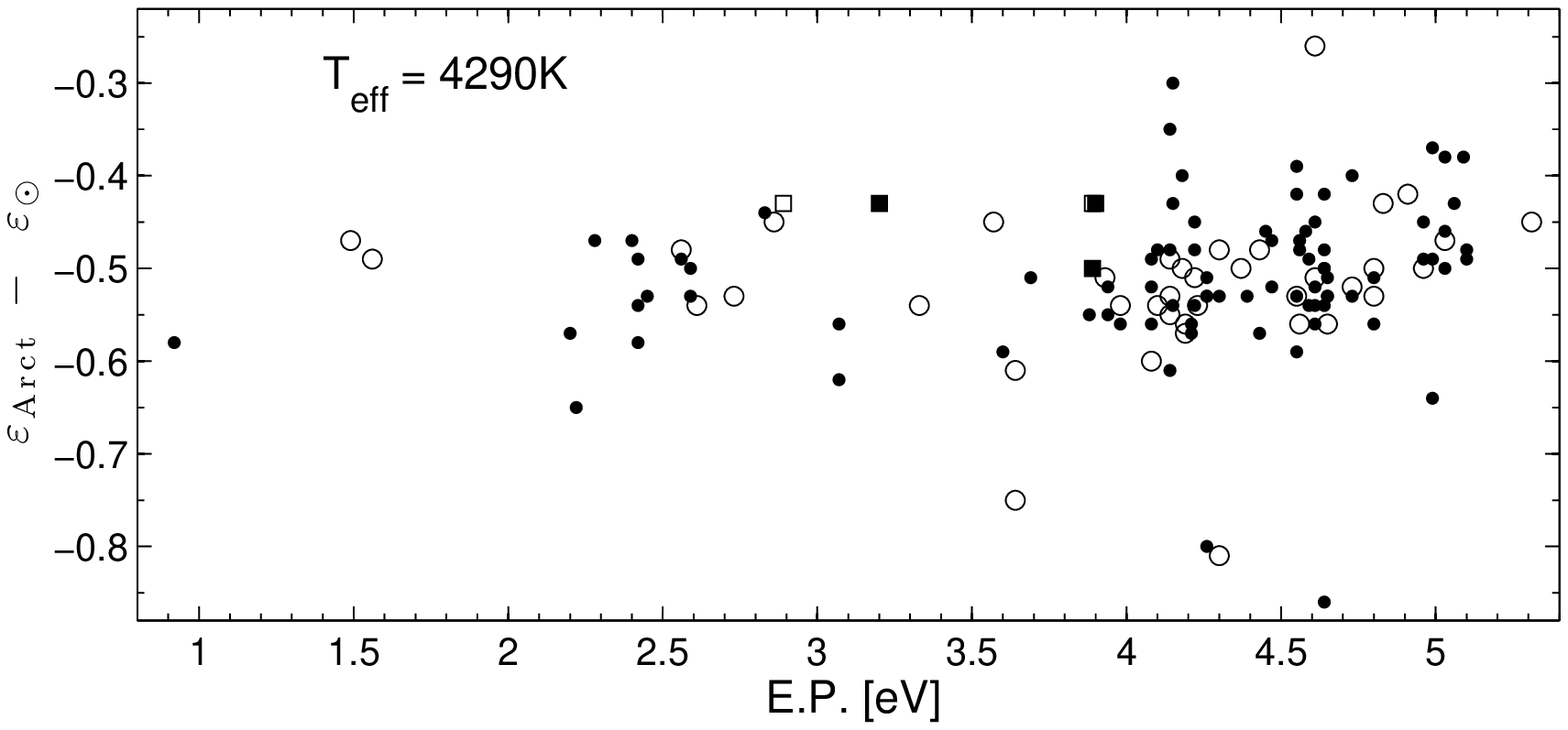}\\
\includegraphics[angle=0,width=0.6\hsize]{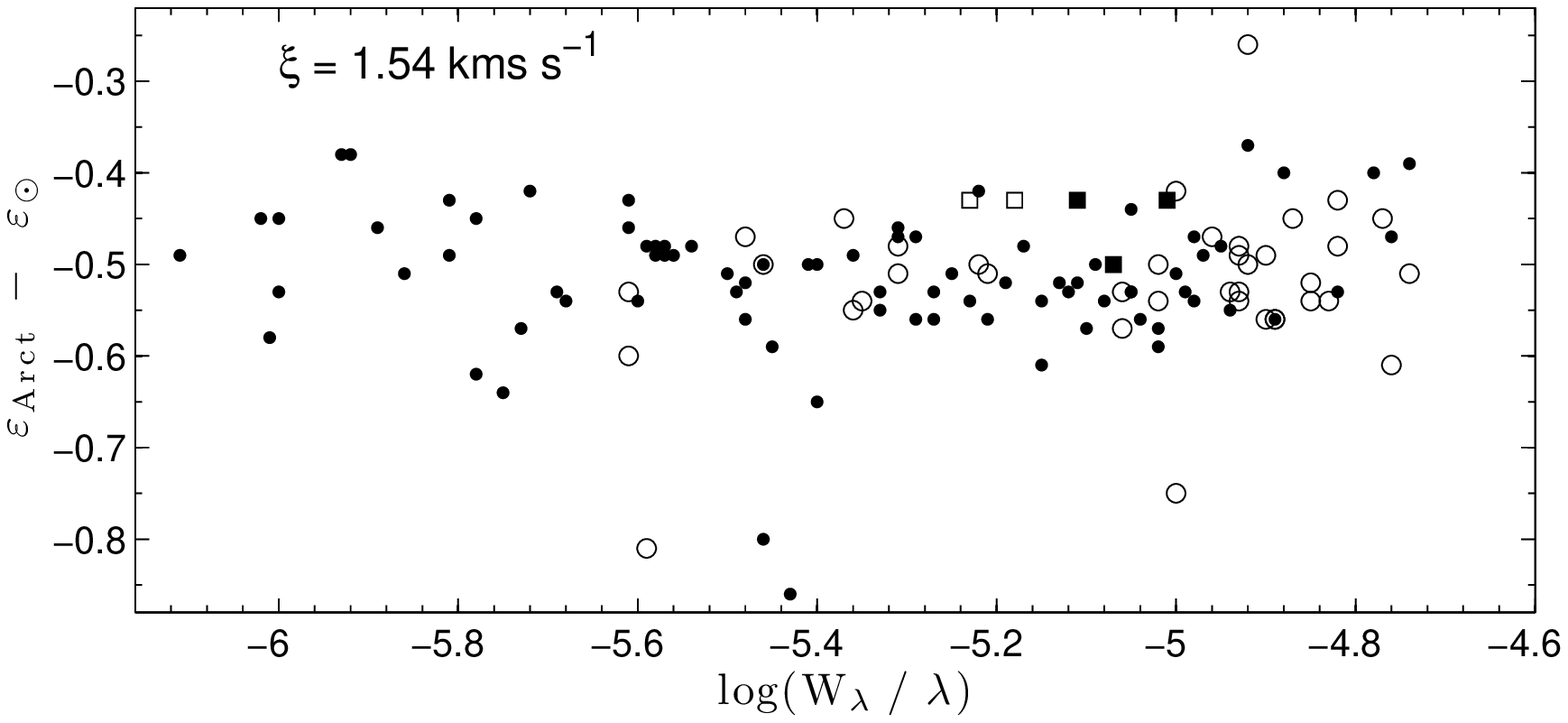}
\end{center}
\caption{Excitation (top panel) and equivalent width (bottom panel) plots for Arcturus relative to the Sun. 
The full linelist employed in this work is indicated as open circles (\ion{Fe}{1}) and open 
squares (\ion{Fe}{2}), while the clean subset from Fulbright et al. (2006a) is shown as filled symbols. The   
few lines that deviate significantly from the mean abundance are likely blends in the solar spectrum.}
\end{figure}
\begin{figure}[!ht]
\includegraphics[angle=0,width=0.5\hsize]{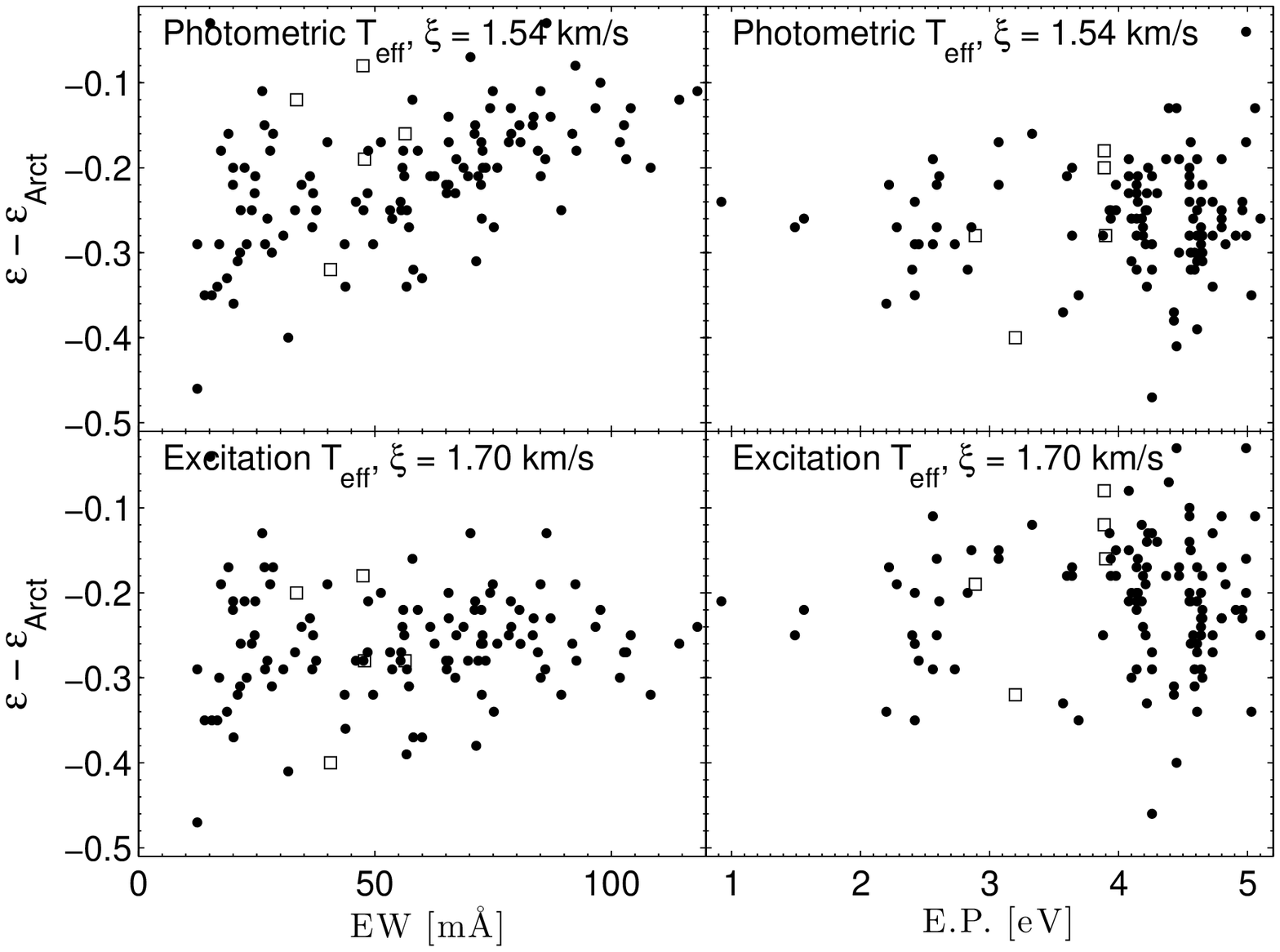}
\includegraphics[angle=0,width=0.5\hsize]{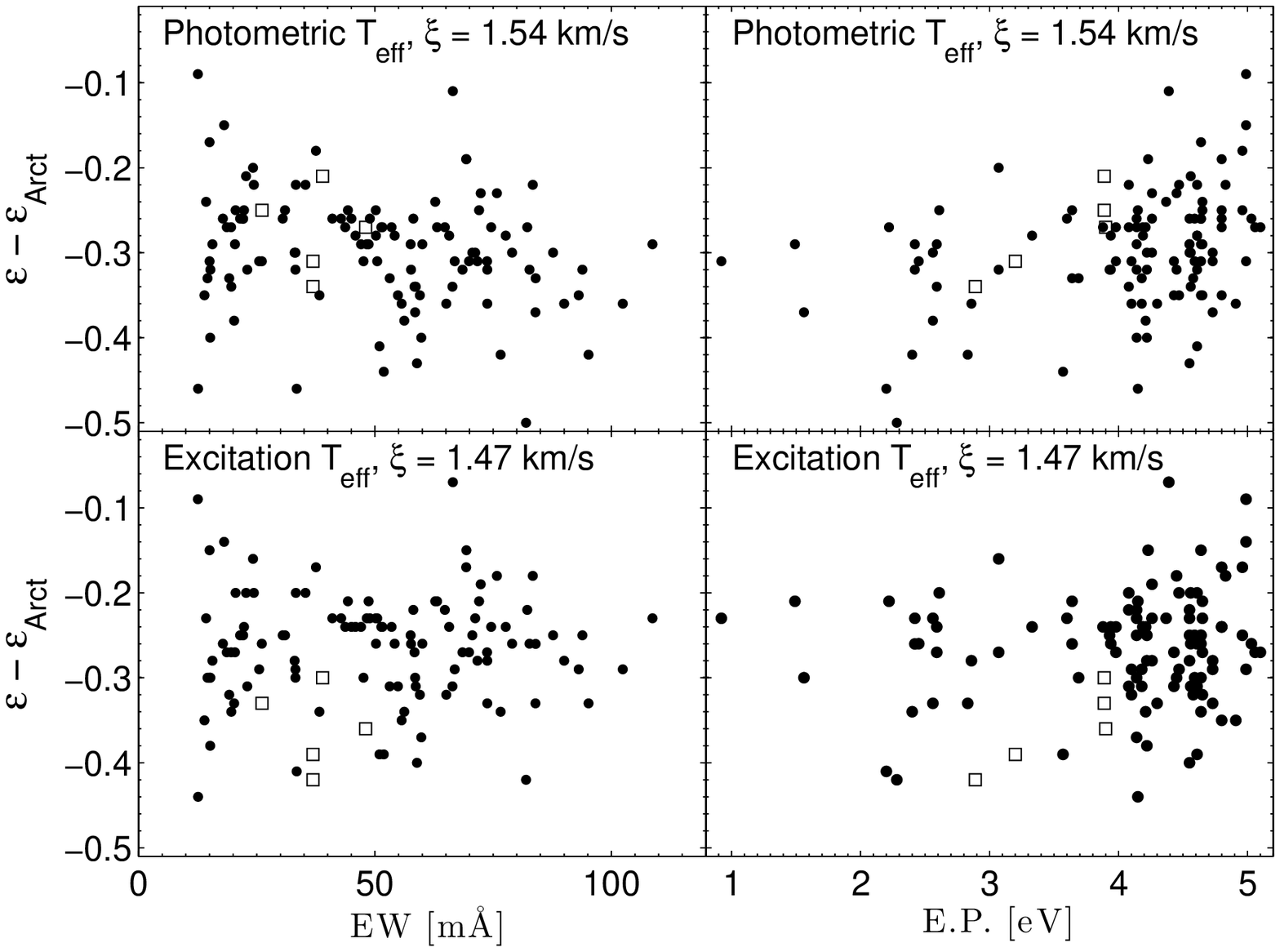}
\caption{Excitation (left panels) and equivalent widths (right panels) plots for the stars \#3 and \#6 
in 47\,Tuc. The top panels show the \ion{Fe}{1} (solid dots) and \ion{Fe}{2} (open squares) 
abundances derived using the photometric temperatures and a  
microturbulent velocity as in Arcturus, while the temperature and microturbulence in the lower 
panels were chosen such as to yield excitation equilibrium and to remove any trend of 
abundance with equivalent widths.}
\end{figure}
\begin{figure}[!ht]
\begin{center}
\includegraphics[angle=0,width=0.7\hsize]{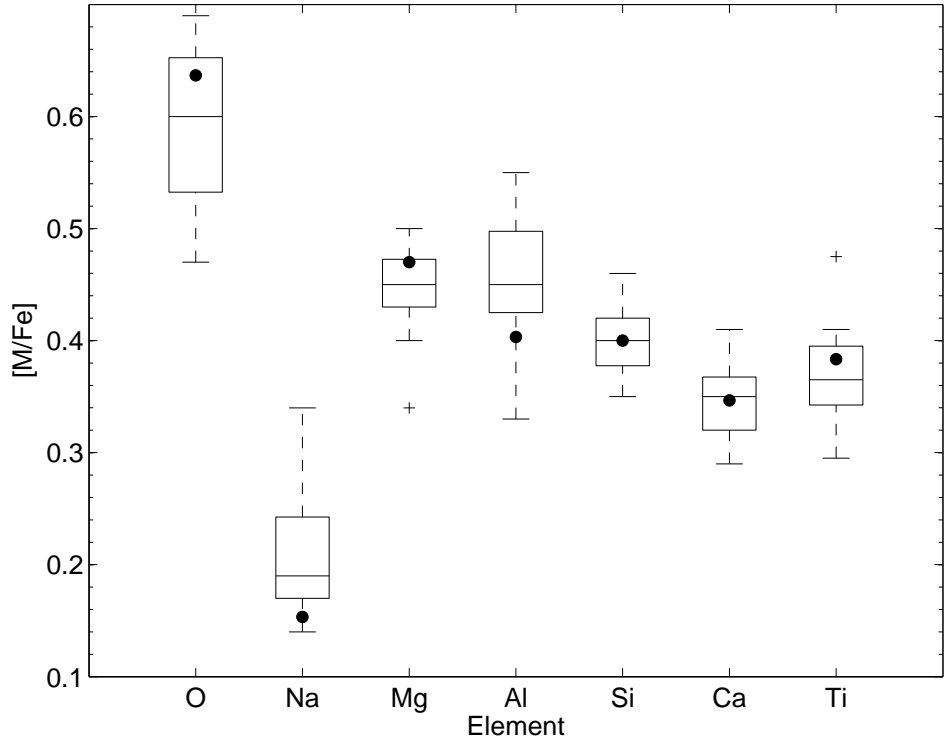}
\end{center}
\caption{Interquartile ranges of our derived abundance ratios. Shown as black dots for comparison 
are the results for Galactic bulge stars at comparable metallicities from F07.}
\end{figure}
\begin{figure}[!ht]
\begin{center}
\includegraphics[angle=0,width=0.7\hsize]{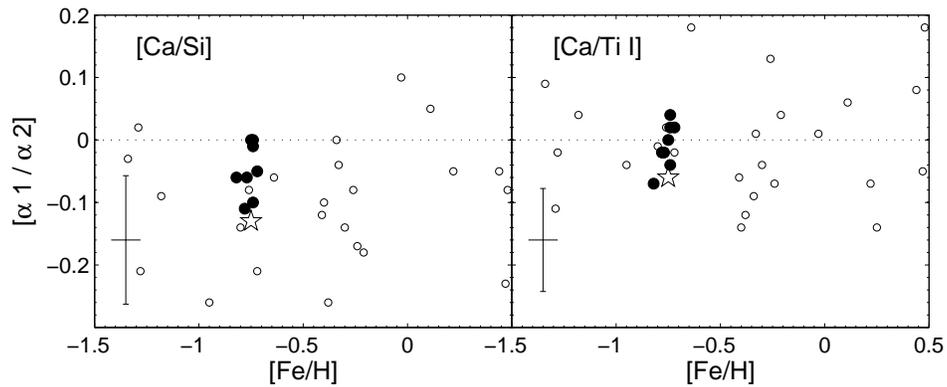}
\end{center}
\caption{A comparison of the [Ca/Si] and [Ca/Ti] abundance ratios 
for the 47\,Tuc stars together with the Galactic bulge sample of F07, shown as open circles. 
A zero point offset of $-$0.09 dex has been applied to the silicon data of F07. Typical 1$\sigma$ 
random errorbars are indicated.}
\end{figure}
\begin{figure}[!ht]
\begin{center}
\includegraphics[angle=0,width=0.7\hsize]{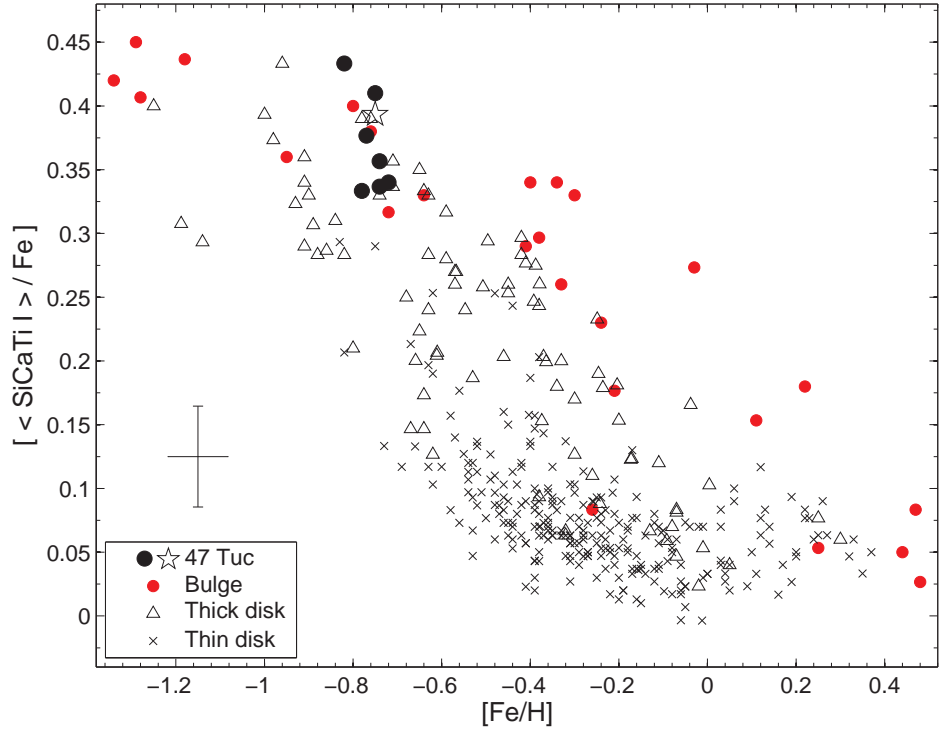}
\end{center}
\caption{A run of the averaged $\alpha$-element abundance ratio vs. iron. The solid circles 
and open star symbol denote the 47\,Tuc red giants and the turnoff star, respectively, 
derived in this work. Additionally shown are observations of Galactic stars from the literature, 
where open circles show the bulge giants of F07, thick disk stars (Prochaska et al. 2000; Fulbright 
2000; Bensby et al. 2005; Brewer \& Carney 2006) are displayed as open triangles and 
thin disk stars (Reddy et al. 2003; Bensby et al. 2005; Brewer \& Carney 2006) are shown 
as crosses. A typical errorbar is indicated towards the lower left.}
\end{figure}
\begin{figure}[!ht]
\begin{center}
\includegraphics[angle=0,width=0.7\hsize]{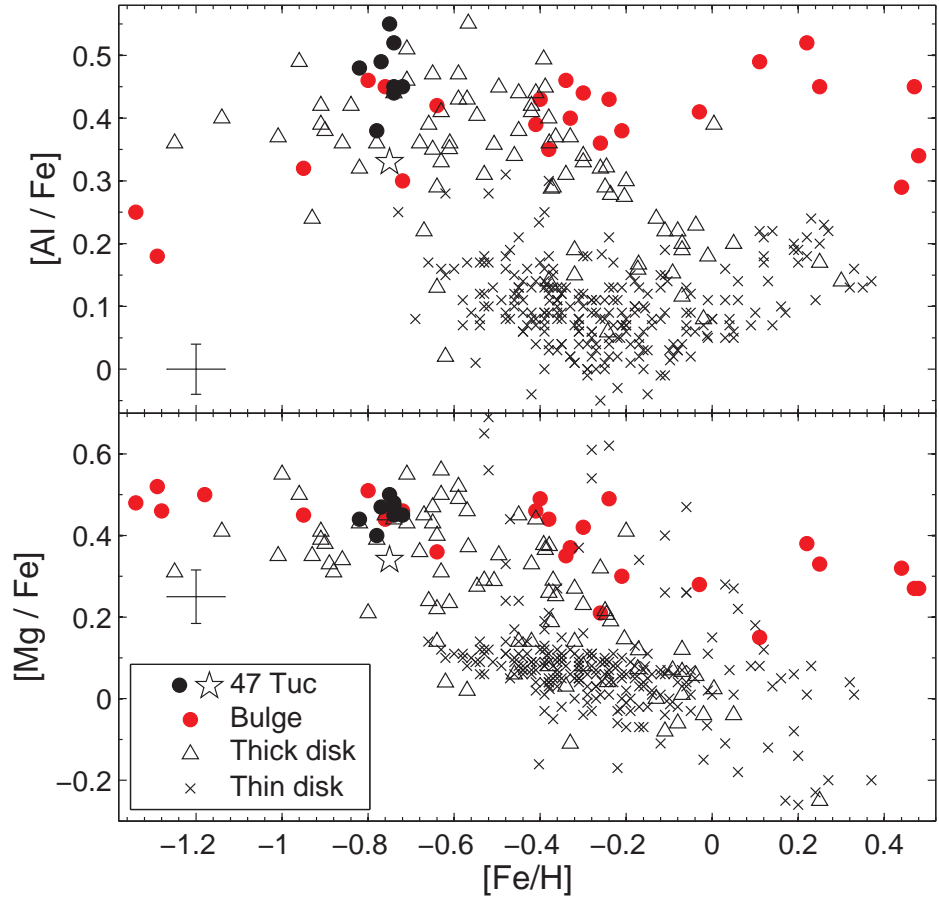}
\end{center}
\caption{Same as Fig.~6, but for the  [Al/Fe] (top panel) and [Mg/Fe] (lower panel) abundance ratios.}
\end{figure}
\begin{figure}[!ht]
\begin{center}
\includegraphics[angle=0,width=0.7\hsize]{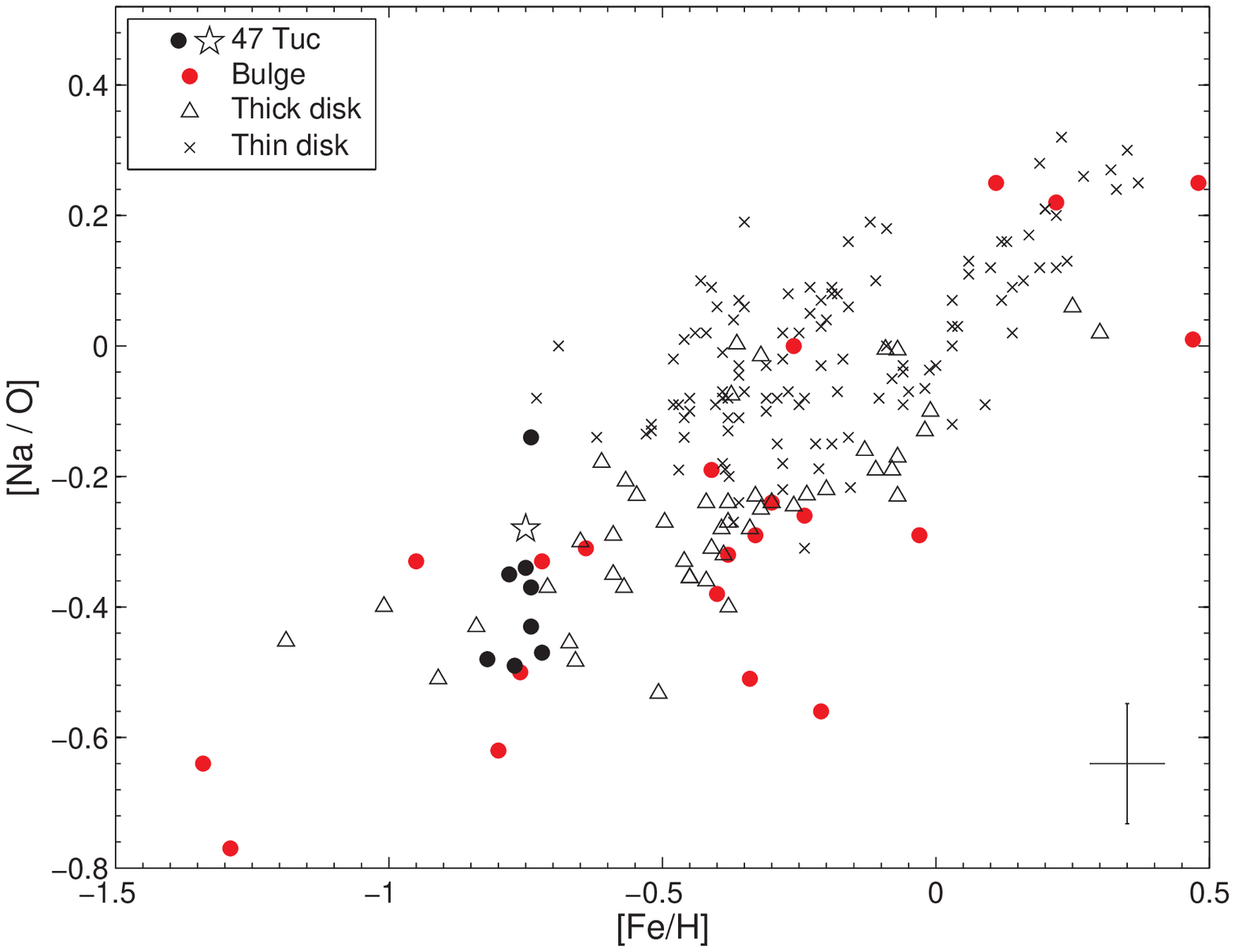}\\
\includegraphics[angle=0,width=0.7\hsize]{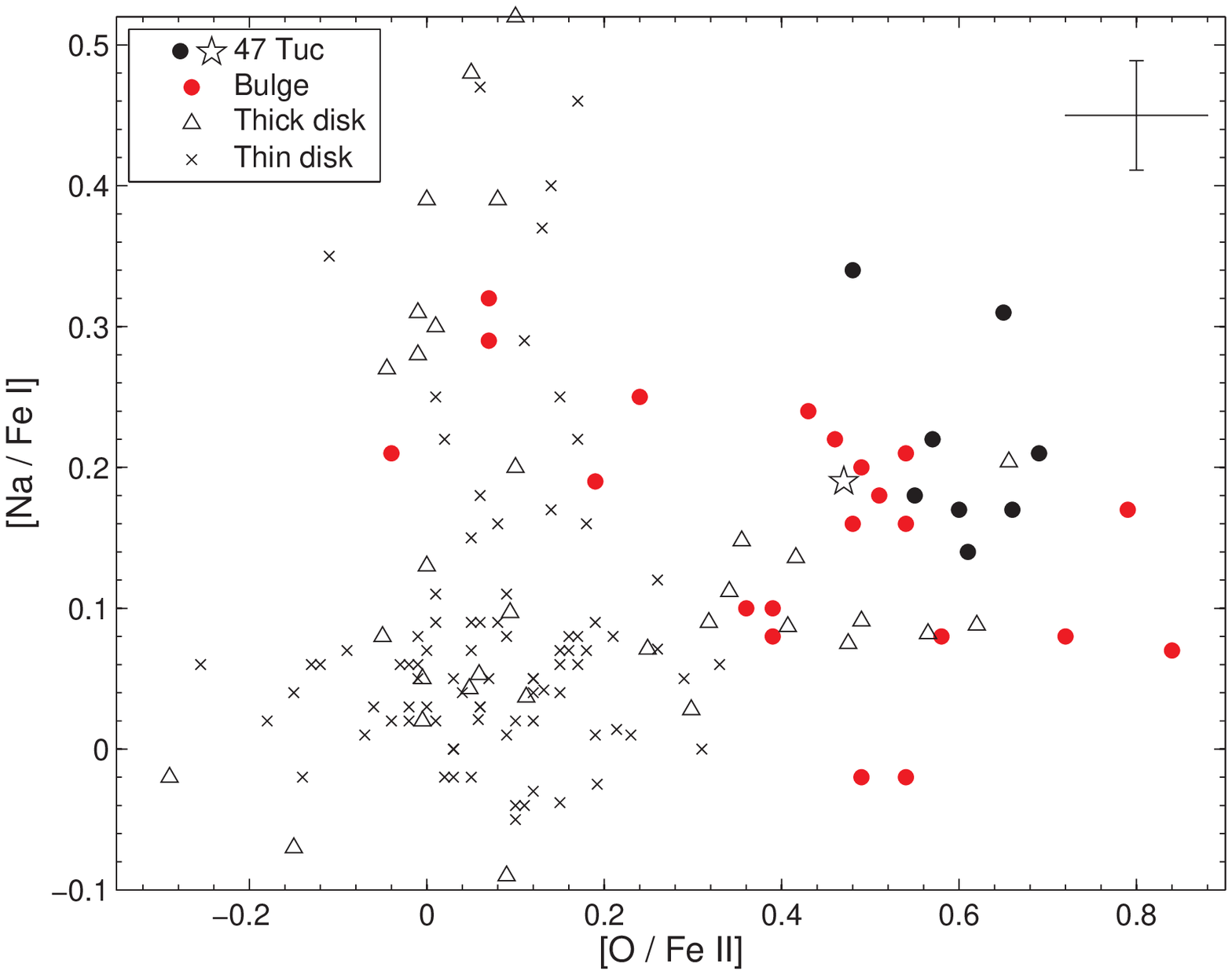}
\end{center}
\caption{Anticorrelation of the sodium and oxygen abundances. The symbols and sources of the data 
are the same as in Figs.~6,7.}
\end{figure}
\clearpage
\begin{table}
\begin{center}
\caption{Log of observations}
\begin{footnotesize}
\begin{tabular}{clcc}
\hline
\hline
&  & {Exposure time} & {S/N} \\
 \raisebox{1.5ex}[-1.5ex]{ID} & \raisebox{1.5ex}[-1.5ex]{Date}  & [s] &  [pixel$^{-1}$] \\
\hline
1 & 2003 Jul 06, 22 & 6060 & 100 \\
2 & 2003 Jul 22, 24 & 3300 & 135 \\
3 & 2003 Jul 24 & 2700 & 150 \\
4 & 2003 Jul 01 & 3600 & 120 \\
5 & 2003 Jul 06 & 3600 & 100 \\
6 & 2003 Jul 22 & 3600 & 120 \\
7 & 2003 Jul 22 & 4500 & 120 \\
8 & 2003 Jul 24 & 3600 & 115 \\
TO\,2 & 2005 Nov 21 & 40000 & 67  \\
Hip\,66815 & 2005 Apr 21 & 120 &  500 \\
\hline
\end{tabular}
\end{footnotesize}
\end{center}
\end{table}
\begin{table}
\begin{center}
\caption{Properties of the target stars}
\begin{footnotesize}
\begin{tabular}{ccccccccr}
\hline
\hline
{ID} & {$\alpha$} & {$\delta$} & {V} & {V$-$I}  & {J$-$K} & {V$-$K} & {\,v$_{r}$} & {$\sigma$\,v$_{r}$} \\
 & \multicolumn{2}{c}{(J2000)} & & & & & \multicolumn{2}{c}{[km\,s$^{-1}$]} \\
\hline
1 & 00 22 07.1 &	 $-$72 00 32.9	 & 12.57 & 1.38 & 0.89 & 3.21 & $-$13.0 & 0.48  \\
2 & 00 22 05.7 &	 $-$72 07 15.7	 & 12.55 & 1.32 & 0.87 & 3.13 & $-$11.7 & 0.49  \\
3 & 00 23 05.2 &	 $-$72 04 28.0	 & 12.53 & 1.29 & 0.80 & 3.00 & $-$17.9 & 0.50  \\
4 & 00 22 32.0 &	 $-$72 06 48.9	 & 12.83 & 1.26 & 0.85 & 3.00 & $-$19.2 & 0.69  \\
5 & 00 22 59.9 &	 $-$72 01 33.6	 & 13.04 & 1.24 & 0.82 & 2.97 & $-$18.6 & 0.65  \\
6 & 00 22 59.7 &	 $-$72 07 06.1	 & 13.18 & 1.26 & 0.81 & 2.97 &  $-$7.5  & 0.51  \\
7 & 00 23 06.3 &	 $-$72 11 10.7	 & 13.32 & 1.22 & 0.78 & 2.85 &  $-$8.2  & 0.45  \\
8 & 00 22 58.1 &	 $-$71 59 32.8	 & 13.50 & 1.21 & 0.78 & 2.81 & $-$26.5 & 0.41  \\
TO\,2 & 00 22 01.5 &   $-$72 10 02.8   & 17.63 & 0.68 &\dots&\dots& $-$15.3 & 0.36 \\
Hip\,66815       &            13 41 42.7 & $-$04 01 45.9 & 8.83 & 0.62 & 0.34 & 1.44 & $-$64.5 & 0.24 \\
\hline
\end{tabular}
\end{footnotesize}
\end{center}
\end{table}
\begin{table}
\begin{center}
\caption{Linelist used for the red giant sample}
\begin{footnotesize}
\begin{tabular}{ccccccccccccc}
\hline
\hline
 {} & {$\lambda$} &  {E.P.} &  \multicolumn{10}{c}{Equivalent Widths [m\AA]} \\
\cline{4-13}
 \raisebox{1.5ex}[-1.5ex]{Ion} & {[\AA]} &  {[eV]} & {Arcturus} & {Sun} & {\#1} & 
{\#2}  & {\#3} & {\#4} & {\#5} & 
{\#6}  & {\#7} & {\#8} \\
\hline
{[}O\,I{]} & 6300.31 & 0.00 &  67.00 & \dots &  71.18 &  62.95 &  67.53 &  58.47 &  47.97 &  53.92 &  49.63 &  44.72  \\
{[}O\,I{]}  & 6363.79 & 0.02 &  30.00 & \dots &  38.00 &  33.25 &  36.32 &  33.26 &  21.80 &  25.39 &  19.93 &  24.92  \\
Na\,I      & 5688.20 & 2.10 & 150.00 & 119.10  & 140.43 & 135.06 & 142.73 & 143.30 & 143.11 & 131.94 & 130.17 & 126.51  \\
Na\,I      & 6154.23 & 2.10 &  73.00 &  40.50  &  60.54 &  55.01 &  60.93 &  68.62 &  68.37 &  53.63 &  51.78 &  45.99  \\
Na\,I      & 6160.75 & 2.10 &  95.00 &  58.20  &  84.29 &  83.27 &  86.86 &  89.03 &  93.17 &  78.27 &  73.58 &  71.45  \\
\hline
\end{tabular}
\end{footnotesize}
\end{center}
\hspace{2.5cm}{\footnotesize Note. --- This Table is published in its entirety in the electronic edition of the {\it Astronomical 
Journal}.\\ \mbox{\hspace{2.5cm}}A portion is shown here for guidance regarding its form and content.}
\end{table}
\begin{table}
\begin{center}
\caption{Linelist used for the turnoff star}
\begin{footnotesize}
\begin{tabular}{cccccc}
\hline
\hline
{} & {$\lambda$} &  {E.P.} &  \multicolumn{3}{c}{Equivalent Widths [m\AA]} \\
\cline{4-6}
 \raisebox{1.5ex}[-1.5ex]{Ion}  & {[\AA]} & {[eV]} & {Hip\,66815} & {Sun} & {TO\,2} \\
\hline
{[}O\,I{]}  & 7774.17   & 9.14 &  44.89 &  63.70  & 51.86 \\
{[}O\,I{]}  & 7775.39   & 9.14 &  36.96 &  53.20  & 37.82 \\
Na\,I       & 5688.20   & 2.10 &  69.14 & 119.10 & 70.61 \\
Na\,I       & 6154.23   & 2.10 &  11.09 &  40.50  &  9.97\\
Na\,I       & 6160.75   & 2.10 &  19.80 &  58.20  & 28.21\\
\hline
\end{tabular}
\end{footnotesize}
\end{center}
\hspace{5cm}{\footnotesize Note. --- This Table is published in its entirety in the \\ \mbox{\hspace{5cm}}electronic edition of the {\it Astronomical 
Journal}. A portion \\ \mbox{\hspace{5cm}}is shown here for guidance regarding its form and content.}
\end{table}
\begin{table}
\begin{center}
\caption{Atmospheric Parameters}
\begin{footnotesize}
\begin{tabular}{cccccc}
\hline
\hline
 {ID} & {T(V$-$K)} & {T(spec)} &   {T(average)} & {log\,$g$}  &
 {$\xi$} \\
 &   [K]      &  [K]       &  [K]  & [cm\,s$^{-2}$] & [km\,s$^{-1}$] \\
\hline
1 &  4191 & 4200 & 4196 & 1.31 & 1.54 \\
2 &  4244 & 4270 & 4257 & 1.36 & 1.54 \\
3 &  4320 & 4310 & 4315 & 1.26 & 1.70 \\
4 &  4321 & 4260 & 4291 & 1.49 & 1.54 \\
5 &  4342 & 4309 & 4326 & 1.60 & 1.54 \\
6 &  4346 & 4380 & 4363 & 1.68 & 1.47 \\
7 &  4431 & 4435 & 4433 & 1.79 & 1.45 \\
8 &  4458 & 4530 & 4494 & 1.90 & 1.45 \\
TO\,2 & \dots & 5750 & \dots & 4.13 & 0.93  \\
Hip\,66815 & 5843 & 5780 & 5812 & 4.41 & 1.13 \\
\hline
\hline
\end{tabular}
\end{footnotesize}
\end{center}
\end{table}
\begin{table}
\begin{center}
\caption{Abundance results for the red giant sample}
\begin{footnotesize}
\begin{tabular}{rrrrrrrrrrrrrrrr}
\hline
\hline
{} & \multicolumn{3}{c}{\#1} & & \multicolumn{3}{c}{\#2} & & \multicolumn{3}{c}{\#3} & & \multicolumn{3}{c}{\#4}\\
\cline{2-4}\cline{6-8}\cline{10-12}\cline{14-16}
 \raisebox{1.5ex}[-1.5ex]{Ion} & {{[}X/Fe{]}} & {$\sigma$}& {N} & & {{[}X/Fe{]}} & {$\sigma$}& {N} & 
              & {{[}X/Fe{]}} & {$\sigma$}& {N} & & {{[}X/Fe{]}} & {$\sigma$}& {N} \\
\hline
{[}Fe\,I/H{]}  & $-$0.77 & 0.06 &    112 & & $-$0.82 & 0.05    & 104 & & $-$0.78 & 0.06    & 109 & & $-$0.75 & 0.07    & 106 \\
{[}Fe\,II/H{]} & $-$0.84 & 0.06 &      4 & & $-$0.95 & 0.07    &   5 & & $-$0.84 & 0.09    &   5 & & $-$0.89 & 0.08    &   5 \\
{[}O\,I{]}$^a$ &    0.66 & \dots &   1 & &    0.69 & \dots &   1 & &    0.57 & \dots &   1 & &    0.65 & \dots &   1 \\
Na\,I          &    0.17 & 0.03 &      3 & &    0.21 & 0.05    &   3 & &    0.22 & 0.02    &   3 & &    0.31 & 0.02    &   3 \\
Mg\,I          &    0.47 & 0.09 &      9 & &    0.44 & 0.04    &   8 & &    0.40 & 0.05    &   9 & &    0.50 & 0.03    &   7 \\
Al\,I          &    0.49 & 0.03 &      3 & &    0.48 & 0.04    &   3 & &    0.38 & 0.05    &   3 & &    0.55 & 0.08    &   5 \\
Si\,I          &    0.41 & 0.08 &     10 & &    0.45 & 0.12    &   9 & &    0.40 & 0.05    &   7 & &    0.41 & 0.11    &   9 \\
Ca\,I          &    0.35 & 0.03 &      9 & &    0.39 & 0.09    &  10 & &    0.29 & 0.03    &   9 & &    0.41 & 0.06    &   9 \\
Ti\,I          &    0.37 & 0.05 &     22 & &    0.46 & 0.05    &  18 & &    0.31 & 0.06    &  16 & &    0.41 & 0.07    &  24 \\
Ti\,II         &    0.32 & 0.12 &      6 & &    0.49 & 0.04    &   4 & &    0.28 & 0.06    &   2 & &    0.37 & 0.10    &   5 \\
\hline
& \multicolumn{3}{c}{\#5} & & \multicolumn{3}{c}{\#6} & &  \multicolumn{3}{c}{\#7}& &  
\multicolumn{3}{c}{\#8} \\
\cline{2-4}\cline{6-8}\cline{10-12}\cline{14-16}
 \raisebox{1.5ex}[-1.5ex]{Ion} & {[}X/Fe{]} & $\sigma$ & N  & & {[}X/Fe{]} & $\sigma$ & N & & {[}X/Fe{]} & $\sigma$ & N & & {[}X/Fe{]} & $\sigma$ & N\\
\hline					      
{[}Fe\,I/H{]}   & $-$0.74 & 0.08   &  76 & & $-$0.74 & 0.05    & 102 & & $-$0.74 & 0.06    & 109 & & $-$0.72 &	 0.05 & 105\\
{[}Fe\,II/H{]}  & $-$0.82 & 0.06   &   4 & & $-$0.80 & 0.05    &   5 & & $-$0.76 & 0.06    &   5 & & $-$0.82 &	 0.04 &   5\\
{[}O\,I{]}$^a$ &    0.48 & \dots &  1 & &    0.60 & \dots &   2 & &    0.55 & \dots &   1 & &    0.61 & \dots &   1\\
Na\,I           &    0.34 & 0.04    &  3 & &    0.17 & 0.02    &   3 & &    0.18 & 0.02    &   3 & &    0.14 &	 0.03 &   3\\
Mg\,I           &    0.48 & 0.06    &  7 & &    0.45 & 0.07    &   9 & &    0.47 & 0.07    &   9 & &    0.45 &	 0.07 &   9\\
Al\,I           &    0.52 & 0.03    &  3 & &    0.45 & 0.06    &   4 & &    0.44 & 0.02    &   3 & &    0.45 &	 0.04 &   3\\
Si\,I           &    0.35 & 0.09    & 10 & &    0.39 & 0.08    &  10 & &    0.37 & 0.08    &  11 & &    0.38 &	 0.08 &   8\\
Ca\,I           &    0.35 & 0.06    & 10 & &    0.29 & 0.09    &   9 & &    0.36 & 0.03    &   9 & &    0.33 &	 0.04 &  10\\
Ti\,I           &    0.31 & 0.07    & 23 & &    0.33 & 0.06    &  17 & &    0.34 & 0.05    &  20 & &    0.31 &	 0.05 &  16\\
Ti\,II          &    0.36 & 0.11    &  6 & &    0.40 & 0.12    &   5 & &    0.37 & 0.05    &   4 & &    0.46 &	 0.06 &   6  \\
\hline
\hline
\end{tabular}
\end{footnotesize}
\end{center}
\hspace{3cm}$^a${The [O/Fe] ratio was taken relative to \ion{Fe}{2} abundances due to similar sensitivity to gravity.}
\end{table}
\begin{table}
\begin{center}
\caption{Abundance results for the turnoff star}
\begin{footnotesize}
\begin{tabular}{rrrr}
\hline
\hline
{} & \multicolumn{3}{c}{TO\,2}\\
\cline{2-4}
 \raisebox{1.5ex}[-1.5ex]{Ion} & {{[}X/Fe{]}} & {$\sigma$}& {N}\\
\hline
{[}Fe\,I/H{]}  & $-$0.75 & 0.13    & 53 \\ 
{[}Fe\,II/H{]} & $-$0.63 & 0.18    &  5 \\
{[}O\,I{]}$^a$     &    0.47 & 0.08    &  2 \\
Na\,I  	       &    0.19 & 0.12    &  2 \\
Mg\,I  	       &    0.34 & 0.11    &  2 \\
Al\,I  	       &    0.33 & 0.01 &  2 \\
Si\,I  	       &    0.46 & \dots    & 1 \\
Ca\,I  	       &    0.33 & 0.08     & 5 \\
Ti\,I  	       &    0.39 & \dots    &  1 \\
Ti\,II 	       &    0.43 & 0.08   &  2 \\
\hline
\hline
\end{tabular}
\end{footnotesize}
\end{center}
\hspace{3cm}$^a${The [O/Fe] ratio was taken relative to \ion{Fe}{2} abundances due to similar sensitivity to gravity.}
\end{table}
\begin{table}
\begin{center}
\caption{Error analysis for the giants \#1 and \#8 and the turnoff star: Given are the deviations of the abundances from 
the values in Tables~6 and 7 upon varying the parameters by the given amount.}
\begin{footnotesize}
\begin{tabular}{ccccccccccc}
\hline
\hline
{} & {} & \multicolumn{2}{c}{$\Delta$T$_{\rm eff}$} & \multicolumn{2}{c}{$\Delta\,\log\,g$} & \multicolumn{2}{c}{$\Delta\xi$} 
& \multicolumn{2}{c}{$\Delta$[M/H]} & {} \\
 \raisebox{1.5ex}[-1.5ex]{\#1} &  \raisebox{1.5ex}[-1.5ex]{Ion}  & {+50\,K}  & {$-$50\,K} & {+0.2 dex} & {$-$0.2 dex} & {+0.1\,km\,s$^{-1}$} & {$-$0.1\,km\,s$^{-1}$} & {+0.1 dex} & {$-$0.1 dex} &  \raisebox{1.5ex}[-1.5ex]{ODF} \\
\hline
& Fe\,I    &   +0.01 & $<$0.01 &   +0.03 & $-$0.06 & $-$0.03 & +0.03 &   +0.01 & $-$0.02 & $-$0.07 \\
& Fe\,II   & $-$0.07 &   +0.09 &   +0.10 & $-$0.16 & $-$0.02 & +0.03 &   +0.04 & $-$0.06 & $-$0.17 \\
&{[}O\,I{]}& $<$0.01 & $-$0.01 &   +0.09 & $-$0.10 & $-$0.01 & +0.01 &   +0.04 & $-$0.04 & $-$0.12 \\
& Na\,I    &   +0.05 & $-$0.04 & $-$0.01 & $<$0.01 & $-$0.03 & +0.04 & $<$0.01 & $<$0.01 & $-$0.02 \\
& Mg\,I    & $<$0.01 & $<$0.01 &   +0.01 & $-$0.03 & $-$0.01 & +0.01 &   +0.01 & $-$0.01 & $-$0.04 \\
& Al\,I    &   +0.03 & $-$0.03 & $-$0.01 & $-$0.01 & $-$0.02 & +0.02 & $-$0.01 & $-$0.01 & $-$0.02 \\
& Si\,I    & $-$0.04 &   +0.05 &   +0.04 & $-$0.08 & $-$0.01 & +0.01 &   +0.02 & $-$0.03 & $-$0.09 \\
& Ca\,I    &   +0.06 & $-$0.05 & $-$0.01 & $-$0.01 & $-$0.05 & +0.05 & $<$0.01 & $<$0.01 & $-$0.04 \\
& Ti\,I    &   +0.07 & $-$0.09 & $<$0.01 & $-$0.01 & $-$0.04 & +0.02 & $-$0.01 & $-$0.01 & $-$0.05 \\
& Ti\,II   & $-$0.02 &   +0.03 &   +0.08 & $-$0.10 & $-$0.03 & +0.05 &   +0.03 & $-$0.03 & $-$0.12 \\
\hline\\
 \raisebox{1.5ex}[-1.5ex]{\#8} &&&&&&&&&&\\
\hline
& Fe\,I    &   +0.02 & $-$0.02 &   +0.02 & $-$0.02 & $-$0.03 & +0.03 &   +0.01 & $-$0.01 & $-$0.06 \\
& Fe\,II   & $-$0.04 &   +0.05 &   +0.09 & $-$0.11 & $-$0.01 & +0.03 &   +0.04 & $-$0.04 & $-$0.16 \\
&{[}O\,I{]}& $-$0.01 &   +0.01 &   +0.08 & $-$0.10 & $-$0.01 & +0.01 &   +0.04 & $-$0.05 & $-$0.12 \\
& Na\,I    &   +0.04 & $-$0.04 & $-$0.01 & $<$0.01 & $-$0.03 & +0.03 & $-$0.01 & $<$0.01 & $-$0.01 \\
& Mg\,I    &   +0.02 & $-$0.01 & $<$0.01 & $<$0.01 & $-$0.01 & +0.01 & $<$0.01 & $<$0.01 & $-$0.02 \\
& Al\,I    &   +0.03 & $-$0.04 & $-$0.01 & $<$0.01 & $-$0.01 & +0.01 & $-$0.01 & $<$0.01 & $-$0.01 \\
& Si\,I    & $-$0.02 &   +0.02 &   +0.03 & $-$0.05 & $-$0.01 & +0.01 &   +0.02 & $-$0.02 & $-$0.06 \\
& Ca\,I    &   +0.05 & $-$0.05 & $-$0.01 & $<$0.01 & $-$0.05 & +0.05 & $-$0.01 &   +0.01 & $-$0.02 \\
& Ti\,I    &   +0.07 & $-$0.08 & $<$0.01 & $<$0.01 & $-$0.03 & +0.03 & $-$0.01 &   +0.01 & $-$0.01 \\
& Ti\,II   & $-$0.01 &   +0.02 &   +0.07 & $-$0.09 & $-$0.04 & +0.03 &   +0.04 & $-$0.03 & $-$0.12 \\
\hline\\
 \raisebox{1.5ex}[-1.5ex]{TO2} &&&&&&&&&&\\
\hline
& Fe\,I    &   +0.07 & $-$0.03 &   +0.07 & $-$0.05 &$<$0.01 &    +0.01 & $<$0.01 & $<$0.01 & $-$0.04 \\
& Fe\,II   & $-$0.01 &   +0.04 & $-$0.03 &   +0.06 &   +0.04 & $-$0.01 &   +0.04 & $<$0.01 & $<$0.01 \\
&{[}O\,I{]}& $-$0.02 &   +0.02 & $-$0.01 &   +0.01 & $-$0.01 & $<$0.01 & $<$0.01 & $-$0.01 & $<$0.01 \\
& Na\,I    & $-$0.01 &   +0.02 & $<$0.01 & $<$0.01 & $<$0.01 & $<$0.01 & $<$0.01 & $<$0.01 &   +0.01 \\
& Mg\,I    & $-$0.01 &   +0.01 & $<$0.01 & $-$0.01 & $-$0.01 & $<$0.01 & $<$0.01 & $-$0.01 & $-$0.01 \\
& Al\,I    & $-$0.04 &   +0.03 & $-$0.06 &   +0.03 & $-$0.03 & $<$0.01 & $<$0.01 & $-$0.02 & $-$0.03 \\
& Si\,I    & $-$0.06 &   +0.04 & $-$0.03 & $<$0.01 & $-$0.03 & $<$0.01 & $-$0.01 & $-$0.01 & $<$0.01 \\
& Ca\,I    & $-$0.01 & $<$0.01 &   +0.06 & $-$0.09 & $-$0.04 &   +0.01 &   +0.01 & $-$0.04 & $-$0.09 \\
& Ti\,I    & $-$0.06 &   +0.04 & $-$0.03 & $<$0.01 & $-$0.03 & $<$0.01 & $-$0.01 & $-$0.01 & $<$0.01 \\
& Ti\,II   & $-$0.01 & $<$0.01 &   +0.06 & $-$0.09 & $-$0.04 &   +0.01 &   +0.01 & $-$0.04 & $-$0.09 \\
\hline
\hline
\end{tabular}
\end{footnotesize}
\end{center}
\end{table}

\end{document}